\documentclass[fleqn,usenatbib]{mnras}

\usepackage{newtxtext,newtxmath}

\usepackage[T1]{fontenc}

\DeclareRobustCommand{\VAN}[3]{#2}
\let\VANthebibliography\thebibliography
\def\thebibliography{\DeclareRobustCommand{\VAN}[3]{##3}\VANthebibliography}

\usepackage{graphicx}	
\usepackage{amsmath}	
\usepackage{orcidlink}
\usepackage[normalem]{ulem}

\title[PSBs dominate low-mass quiescent population]{\textit{JWST} PRIMER: Post-starburst galaxies dominate the low-mass quiescent galaxy population since $z \sim 2$}

\author[T. de Lisle et al.]
{Thomas de Lisle\textsuperscript{\,\orcidlink{0009-0004-1726-0260}},$^{1}$\thanks{E-mail: ppytdd@nottingham.ac.uk}
David~T.~Maltby\textsuperscript{\,\orcidlink{0000-0002-8163-080X}},$^{1}$
Omar~Almaini\textsuperscript{\,\orcidlink{0000-0001-9328-3991}},$^{1}$
Vivienne~Wild\textsuperscript{\,\orcidlink{0000-0002-8956-7024}},$^{2}$
Elizabeth~Taylor\textsuperscript{\,\orcidlink{0000-0001-8728-2984}},$^{3}$
\newauthor Kate~Rowlands\textsuperscript{\,\orcidlink{0000-0001-7883-8434}},$^{4,5}$
Guillaume~Hewitt\textsuperscript{\,\orcidlink{0009-0006-7827-007X}},$^{1}$
Maya Skarbinski\textsuperscript{\,\orcidlink{0009-0004-0844-0657}},$^{4}$
James~S.~Dunlop\textsuperscript{\,\orcidlink{0000-0002-1404-5950}},$^{3}$
Adam~C.~Carnall\textsuperscript{\,\orcidlink{0000-0002-1482-5818}},$^{3}$
\newauthor Anton~M.~Koekemoer\textsuperscript{\,\orcidlink{0000-0002-6610-2048}},$^{6}$
Norman A. Grogin\textsuperscript{\,\orcidlink{0000-0001-9440-8872}},$^{6}$
and Derek~J.~McLeod\textsuperscript{\,\orcidlink{0000-0003-4368-3326}}$^{3}$\\
$^{1}$School of Physics and Astronomy, University of Nottingham, University Park, Nottingham NG7 2RD, UK\\
$^{2}$School of Physics and Astronomy, University of St Andrews, North Haugh, St Andrews KY16 9SS, UK\\
$^{3}$Institute for Astronomy, University of Edinburgh, Royal Observatory, Edinburgh EH9 3HJ, UK\\
$^{4}$William~H.~Miller III Department of Physics and Astronomy, Johns Hopkins University, Baltimore, MD 21218, USA\\
$^{5}$AURA for ESA, Space Telescope Science Institute, 3700 San Martin Drive, Baltimore, MD 21218, USA\\
$^{6}$Space Telescope Science Institute, 3700 San Martin Drive, Baltimore, MD 21218, USA}

\date{Accepted 2026 July 13. Received 2026 June 16; in original form 2026 February 11}

\pubyear{\the\year{}}

\begin{document}
\label{firstpage}
\pagerange{\pageref{firstpage}--\pageref{lastpage}}
\maketitle

\begin{abstract}
We present the evolution of the stellar mass functions for photometrically-selected star-forming, passive and post-starburst galaxies, based on \textit{JWST} PRIMER observations of the Ultra-Deep Survey (UDS) field. We identify over 800 post-starburst galaxies within the redshift range $0.5 < z < 3.0$. We confirm the presence of a low-mass upturn in the quenched galaxy mass function at $\log(M_{\ast}/\mathrm{M_\odot}) < 10$ out to $z\sim2$, and show it is primarily driven by post-starburst galaxies. These findings imply that a rapid quenching pathway for low-mass galaxies is already active by $z \sim 2$, which may be linked to environmental processes. If we assume that all post-starburst galaxies evolve into passive galaxies, adopting a typical visibility time for the post-starburst phase, we find a significant overprediction of low-mass passive galaxies at later times. We suggest a variety of factors that could contribute to reconciling this tension. If low-mass ($M_{\ast} \lesssim 10^{10}\,\mathrm{M_{\odot}}$) post-starburst galaxies quench through slower channels than high-mass systems then the post-starburst phase may be visible for longer periods, reducing their contribution to the passive population. However, the maximum realistic visibility timescale cannot account for all of the excess. We propose that additional factors inhibit the build-up of the low-mass passive galaxy mass function. 
These galaxies may be removed through mergers or tidal disruption, rejuvenate and return to the star-forming population, or represent systems observed during a dormant phase of stochastic star formation. We conclude that the emergence of the low-mass quenched population is likely governed by a complex interplay of processes, and cannot be explained by simple one-way evolutionary pathways.
\end{abstract}




\section{Introduction}

Galaxies in the local Universe show a clear bimodality in their morphologies and spectral types. Typically, high-mass galaxies are red, quiescent and spheroidal, whereas less massive galaxies are blue, star-forming and disc-dominated \citep[e.g.][]{strateva_color_2001, baldry_quantifying_2004, taylor_galaxy_2015}. These two groups are known as the red sequence and blue cloud, respectively. The precise origin of this bimodality is still an area of ongoing research. It has been found that the number density and stellar-mass density of the passive galaxy population has been continuously increasing since $z = 3$--$4$ \citep[e.g.][]{bell_nearly_2004, faber_galaxy_2007, muzzin_evolution_2013}. As rejuvenation of passive galaxies is thought to be low \citep[e.g.][]{treu_assembly_2005, thomas_environment_2010, belli_kmos3d_2017}, the primary mechanism for this build-up of mass on the red sequence is believed to be through the quenching of star-forming galaxies. This would cause galaxies to transition from the blue cloud to the red sequence. In order to account for the structural differences between galaxies in these two populations, it is thought that galaxies also undergo a morphological transition either shortly before or during the quenching of star-formation \citep[e.g.][]{whitaker_large_2012, almaini_massive_2017}.

Many possible quenching mechanisms have been proposed to account for the observed evolution, and these can be broadly separated into two categories: mass quenching and environmental quenching. The mechanisms that fall under mass quenching are related to star-formation rate (SFR), and hence the stellar mass of the galaxy \citep{peng_mass_2010}. These include the heating or expulsion of gas from a galaxy due to either stellar or AGN feedback \citep[e.g.][]{hopkins_black_2005, bower_dark_2017, donnari_quenched_2021}. It has also been proposed that compact spheroidal galaxies may be produced from disc instabilities and contraction due to cold gas inflow \citep[e.g.][]{dekel_formation_2009}. In contrast, environmental quenching is comprised of mechanisms that are caused by a galaxy's environment, rather than its mass.  These include ram-pressure stripping of the interstellar medium \citep[ISM; e.g.][]{gunn_infall_1972}, strangulation \citep{larson_evolution_1980}, galaxy harassment \citep{farouki_computer_1981}, and mergers \citep{toomre_mergers_1977}.  Low-mass galaxies are believed to be primarily quenched through these environmental processes, whereas high-mass galaxies are largely expected to be quenched through mass-dependent mechanisms \citep[e.g.][]{peng_mass_2010}.

To investigate galaxy quenching it is useful to study a class of transition galaxy known as post-starburst (PSB) galaxies, which have been rapidly quenched within the past Gyr following a burst of star-formation. Spectroscopically, they are identified through the presence of strong Balmer absorption lines in galaxies showing little ongoing star-formation, which suggests that they contain an increased proportion of A-type stars \citep{dressler_spectroscopy_1983, wild_post-starburst_2009}. Galaxies with these spectral features are likely to have experienced a burst of star-formation before being rapidly quenched, if the features are strong enough \citep{wild_post-starburst_2009}. However, weaker spectral features may still indicate rapid quenching, without the need for a preceding starburst.

Due to their transitional nature, PSBs are intrinsically rare, for example only $\sim1$ per cent of local galaxies in the Sloan Digital Sky Survey (SDSS) appear to be PSBs \citep{goto_266_2005}. There are signs that a significant fraction of those observed locally are the result of merger events \citep[e.g.][]{zabludoff_environment_1996, blake_2df_2004, goto_266_2005, yang_detailed_2008, pawlik_shape_2016, wilkinson_merger_2022, ellison_galaxy_2022}, and there is also evidence to suggest that they have significant residual cold gas reservoirs \citep[e.g.][]{zwaan_cold_2013, rowlands_evolution_2015, french_discovery_2015, ellison_low_2025}. The combination of these findings may imply that low-redshift star-forming galaxies quenching through mergers may require multiple merger events for their star-formation to fully quench \citep{rowlands_evolution_2015}.

Until recently, few PSBs were spectroscopically identified at $z > 1$ \citep[e.g.][]{vergani_kgalaxies_2010, bezanson_massive_2013, williams_morphology_2017}, which led to the development of photometric methods for identifying PSBs at high redshift.  \cite{wild_new_2014} applied a Principal Component Analysis (PCA) to deep multiwavelength photometry taken in the UDS field to identify large samples of PSBs. This photometric method has since been verified with deep optical spectroscopy from the VLT \citep{maltby_identification_2016}.

PSBs have been shown to be generally of low mass in the redshift range $0.5 < z < 1$, mostly with disc-like morphologies \citep{maltby_structure_2018}. However at $z > 1$ a higher mass population becomes more prevalent, which are typically compact and spheroidal \citep{whitaker_large_2012, almaini_massive_2017, maltby_structure_2018, wu_fast_2018}. The exceptionally compact nature of high-mass PSBs at $z>1$ suggests they formed through a gas-rich compaction event, following a merger or disc collapse. \cite{maltby_high-velocity_2019} also discovered that massive PSBs at $z > 1$ have high-velocity outflows, which also appear to be present up to $\sim 1$ Gyr after quenching \citep{Taylor2024}, suggesting recurring AGN activity as a likely driver. In general, however, massive ($M_{\ast}$ > $10^{10.5}\,\mathrm{M_{\odot}}$) PSBs in the redshift range $1<z<3$ show no evidence for excess X-ray AGN activity compared to older passive galaxies \citep{almaini_no_2025}, suggesting that any AGN activity is short-lived and stochastic, and perhaps not directly linked to the primary quenching event.

Regarding low-mass PSBs, an excess has been identified in cluster environments out to at least $z\sim1$ \citep[e.g.][]{muzzin_phase_2014, socolovsky_enhancement_2018, mcnab_gogreen_2021}. Clustering analyses also reveal that these low-mass PSBs are strongly clustered, consistent with being satellite galaxies residing in very massive dark matter haloes \citep{wilkinson_starburst_2021}. The fact that these PSBs exhibit disc-dominated morphologies \citep{maltby_structure_2018}, suggests a relatively gentle quenching mechanism that did not significantly disrupt the rotationally supported system, and one potentially linked to the cluster environment (e.g. gas stripping). These findings suggest that high- and low-mass PSB populations are typically quenched through different mechanisms; those at low-mass generally experience environmental quenching, and those at high-mass are internally quenched.

It is unclear when environmental processes begin to play a significant role in the quenching of galaxies.  Based on numerical models, environmental quenching of low-mass galaxies is expected in massive halos, though it is unclear if these processes provide a significant quenching route at very early epochs.  For example, \cite{donnari_quenched_2021} found that the IllustrisTNG simulation predicts that the majority of quenched low-mass galaxies are satellites of groups or clusters. This suggests environmental quenching, though it primarily acts below $z\sim2$. However, protoclusters have been discovered at $z > 2$ \citep[e.g.][]{ata_predicted_2022}, and there is tentative evidence of a relative excess of low-mass passive galaxies ($M_\ast < 10^{10}\,\mathrm{M_{\odot}}$) in these protoclusters compared to the field in the redshift range $2 < z < 2.5$ \citep{edward_stellar_2024}.  Furthermore, there is evidence that the environment may contribute to the quenching of galaxies much earlier, though indirectly. For example, the IllustrisTNG simulations predict that galaxies at $z > 4$ residing in overdense regions are more likely to quench.  This is due to the higher availability of infalling gas, which fuels early SMBH growth, and hence quenching through AGN feedback \citep{kurinchi-vendhan_origin_2024}.  Therefore a correlation of quenching with environment may not necessarily imply a direct causative link.

Studying the galaxy stellar-mass functions (GSMFs) of different populations can provide some insight into the nature of quenching. Mass functions for star-forming samples in the local Universe have been found to be adequately fit by a single Schechter function \citep[e.g.][]{baldry_galaxy_2012}, and the evolution of the star-forming GSMF has been shown to be relatively minor out to at least $z > 2$ \citep[e.g.][]{tomczak_galaxy_2014}. If quenching is dominated by mass-driven mechanisms, then the mass functions of passive populations should be well explained by a single Schecter function only with a higher characteristic mass than that of the best-fitting Schecter function for star-forming populations. However, it has been found that passive GSMFs are usually best fit with double Schecter functions out to about $z = 1.5$ \citep[e.g.][]{mcleod_evolution_2021, gully_insights_2025}, which is believed to be further evidence for two mass-dependent quenching routes, and hence environmental quenching. Evidence has since been found that this upturn in the low-mass end of the passive GSMF may be present as early as $z\simeq2.25$ \citep{santini_stellar_2022, hamadouche_jwst_2025} or even $z\simeq3.25$ \citep{shuntov_stellar_2026}.

Investigating recently quenched galaxies and older passive galaxies separately can provide further insight into the initial quenching mechanisms and the longer-term evolution of quiescent galaxies. The identification of large samples of PSBs at high redshift using  photometric techniques led to the first study of the evolution of the PSB galaxy stellar-mass function between $z = 0.5$ and $z = 2.0$ in \cite{wild_evolution_2016}. They found that the shape and normalisation of the PSB mass function evolves rapidly. At high redshift, the PSB mass function is comparable in shape to that of passive galaxies, while at lower redshifts ($z<1$) an upturn is observed at lower stellar masses ($M_{\ast}<10^{10}\,\mathrm{M_{\odot}}$), which was attributed to the likely emergence of mass-independent environmental quenching. This work was extended by \cite{taylor_role_2023} who used ground-based data from the UDS survey to study PSB mass functions, separated by environment, and found evidence for a low-mass upturn in the abundance of PSBs in dense environments out to $z = 1.5$. The low-mass upturn found by \cite{taylor_role_2023} was found to occur at $M_{\ast}<10^{10}\,\mathrm{M_{\odot}}$, but deeper data is required to investigate the presence of this feature at higher redshift.

In this work, we present the first results from an application of the photometric PCA technique \citep{wild_new_2014} to the \textit{JWST} observations from the Public Release IMaging for Extragalactic Research (PRIMER) program.  These observations allow us to classify galaxies at much lower stellar masses and to higher redshifts than previously possible.  We present the first mass functions of the post-starburst population that extend to $\log(M_\ast/\mathrm{M_\odot}) < 10$ above $z = 2$, enabling the search for evidence of environmental quenching at this epoch. We then estimate the contribution of PSBs to the build-up of the quiescent galaxy mass function, to determine the fraction of passive galaxies that pass through a PSB phase.  We adopt a cosmology with $\Omega _{\rm m} = 0.3$, $\Omega _{\Lambda} = 0.7$ and $h = 0.7$.  All magnitudes are given in the AB system.

\section{Data}
This section describes the data that we use in this work: The \textit{JWST} observations in the near-infrared from the Public Release IMaging for Extragalactic Research (PRIMER; PI:Dunlop) project, along with auxiliary imaging in three {\em HST} bands from the Cosmic Assembly Near-infrared Deep Extragalactic Legacy Survey \citep[CANDELS;][]{grogin_candels_2011, koekemoer_candels_2011}, and the UKIRT Infrared Deep Sky Survey (UKIDSS; \citealt{lawrence_ukirt_2007}) Ultra Deep Survey (UDS; PI:Almaini) Data Release 11 (DR11). We perform our analysis with the space-based data from PRIMER and CANDELS, however, we use overlapping data and results from the ground-based UDS as a comparison study, the details of which are in Section~\ref{section:UDS Comparison}.

\subsection{PRIMER observations with \textit{JWST}}
The PRIMER project \citep{dunlop_primer_2021} is a large public legacy survey that was awarded 190 hours on \textit{JWST}, and has taken observations of the UDS and COSMOS fields (covering 234 and 144 sq. arcmin, respectively).  This used the NIRCam and MIRI instruments, providing coverage in the near-infrared (NIR) and mid-infrared (MIR), respectively. In this work we use the PRIMER v1.0 data release of the 8-band \textit{JWST} NIRCam observations ($F090W$, $F115W$, $F150W$, $F200W$, $F277W$, $F356W$, $F410M$, $F444W$) within the PRIMER-UDS field, providing wavelength coverage from $0.9$--$4.4\,\mu\mathrm{m}$.

\subsubsection{Auxillary HST Imaging}
To probe shorter observed-frame wavelengths, we use imaging observations from three {\em HST}/ACS bands ($F435W$, $F606W$, $F814W$), ranging from $0.4\,\mu\mathrm{m}$ to $0.8\,\mu\mathrm{m}$, covering the PRIMER-UDS field from CANDELS \citep{grogin_candels_2011, koekemoer_candels_2011}. \cite{stevenson_primer_2026} finds that the combination of both optical {\em HST} and \textit{JWST} NIRCam imaging is required to more accurately identify massive $M_{\ast} > 10^{10}\rm\,M_{\odot}$ quiescent galaxies at $z > 2$. Furthermore, as photometry is required below $4000$\,\AA\ in the rest-frame to accurately classify galaxies using the PCA method \citep{wild_new_2014}, the inclusion of these {\em HST}/ACS observations allows us to extend our analysis down to $z = 0.5$, whereas only using NIRCam observations would restrict our analysis to above $z \sim 1.5$.

\subsubsection{PRIMER Source Detection and Photometric Redshifts}

We perform source detection in the PRIMER-UDS field via {\sc SExtractor}, using $F200W$ as the detection band. Aperture photometry is extracted in $0.5\arcsec$ apertures from each of the 11 available wavebands, centred at the position of the $F200W$ detections. The $5\mathrm{\sigma}$ depth of this extraction is $\simeq 27.5$ in $F200W$ based on the variance between apertures placed in source-free regions, and is consistent across all bands ($\simeq 27$ - $28$). However, we use a conservative magnitude limit of $26.5$ in the $F200W$ $0.5\arcsec$ fixed-aperture magnitudes for our analysis to allow for depth variations across the PRIMER-UDS field, and to ensure the galaxies have sufficiently high signal-to-noise in their SEDs for robust classifications. We also use these $0.5\arcsec$ fixed-aperture magnitudes for the purposes of calculating the mass completeness limits of our samples, described in Section~\ref{section:Sample Selection}. Errors in the photometry were determined by combining errors in the source counts (set at $5\%$ of the source flux) with errors in the background. Background errors were determined for each
source using the variance of approximately 200 nearby pure sky regions.

To correct for aperture effects on the extracted fluxes, we first apply a wavelength-dependent PSF correction determined from empirical PSFs on a band-by-band basis. These PSFs were generated by stacking $\sim$100 bright unsaturated stars in each waveband. The resultant PSF-corrected magnitudes are then corrected to total, using an indicative correction based on that required for the $F200W$ detection band. This is determined from the difference between the $F200W$ PSF-corrected and {\sc SExtractor} total magnitude ($\texttt{magbest}$). For each source, this second correction only affects the normalisation of the object SED, and not the overall shape (i.e.\ colours). The initial multiwavelength extraction catalogue contains 133\,172 sources.  We first mask objects with potentially compromised photometry, due to their proximity to diffraction spikes and detector edges. Stars are also identified and removed from our catalogues. This is achieved via comparing the extracted photometry in two `uncorrected' apertures ($0.35\arcsec$ and $2\arcsec$), and the identification of a stellar locus. This leaves $127\,617$ galaxies. We also remove stellar-like objects with {\sc SExtractor} $\texttt{class\_star} > 0.9$, which leaves $117\,100$ galaxies. Finally, we apply a magnitude cut of $F200W$ $< 26.5$, significantly reducing the number of galaxies to $36\,855$.

Photometric redshifts are determined using the {\sc eazy} software \citep{brammer_eazy_2008}.  Here, the 11-band photometry is fit using a grid of galaxy templates spanning a wide range of ages, and with the addition of dust-reddened SEDs.  We compare our photometric redshifts to $\sim800$ high-quality spectroscopic redshifts available in the PRIMER-UDS field (see Section~\ref{section:Spectroscopic Data}, for further details). The uncertainty in the distribution of $\frac{\left|z_{\rm phot}-z_{\rm spec}\right|}{(1+z_{\rm spec})}$ is found to be $\sigma_{\textrm{NMAD}} \sim 0.019$ over the redshift range $0 < z < 5$, with an outlier fraction [$\frac{\left|z_{\rm phot}-z_{\rm spec}\right|}{(1+z_{\rm spec})} > 0.15$] of $<5$ per cent, where $\sigma_{\textrm{NMAD}}$ is the normalised median absolute deviation.  We limit our sample to the redshift range $0.5 < z < 3.0$ in order to ensure the robustness of our classifications, and due to our passive sample becoming incomplete below $M_{\ast} \approx 10^{10}\,\mathrm{M_{\odot}}$ at $z = 3.0$ (see Section~\ref{section:Sample Selection} for details). This leaves $29\,970$ galaxies. $6782$ of these objects have incomplete data in the required rest-frame wavelength range for our analysis due to the incomplete overlap between the {\em HST}/ACS and NIRCam observations, leaving a final catalogue of $23\,188$ galaxies that we input into our analysis which is described in Section~\ref{section:SC Analysis}.

\subsection{UKIDSS Ultra-Deep Survey}
The UKIDSS \citep{lawrence_ukirt_2007} UDS (PI: Almaini), is a deep, large-scale near-infrared (NIR) ground-based survey covering 0.77\,deg\textsuperscript{2}. The latest release (UDS DR11) is currently the deepest $K$-band survey over such a large area and contains \textit{JHK} photometry to limiting depths of $J = 25.6$, $H = 25.1$, and $K = 25.3$ (AB; $5\mathrm{\sigma}$ in 2 arcsec apertures). There are also complementary multiwavelength observations. The additional imaging includes optical $BVRi'z'$ photometry from the Subaru--\emph{XMM–Newton} Deep Survey \citep{furusawa_subaruxmm-newton_2008}, mid-infrared ($3.6\,\mathrm{\mu m}$ and $4.5\,\mathrm{\mu m}$) from the \emph{Spitzer} UDS Legacy Program (SpUDS; PI: Dunlop), $u'$-band observations from MegaCam on the Canada--France--Hawaii Telescope (CFHT), and $Y$-band photometry from the VISTA VIDEO survey (\citealt{jarvis_vista_2013}). The 12-band multiwavelength catalogue based on these data was presented in \cite{wilkinson_starburst_2021}, along with PCA-based galaxy classifications. We compare this catalogue to our PRIMER-based catalogue and classifications in Section~\ref{section:UDS Comparison}.

\subsection{Spectroscopic Data}
\label{section:Spectroscopic Data}
Due to a variety of spectroscopic observations covering the PRIMER-UDS field, we have a large sample of galaxies for which spectroscopic classifications are available, and we compare these with our photometric classifications in Section~\ref{section:Spectroscopic Comparison}. Part of our spectroscopic data comes from The Early eXtragalactic Continuum and Emission Line Survey (EXCELS; GO 3543; PIs: Carnall, Cullen; \citealt{carnall_JWST_2024}), a NIRSpec complement to PRIMER, that observed a sample of $\sim 100$ galaxies at cosmic noon. We also use a combination of spectroscopy from the VANDELS project (ESO programme 194.A-2003; \citealt{mclure_vandels_2018}; \citealt{pentericci_vandels_2018}), the UDSz project \citep{bradshaw_high-velocity_2013, mclure_sizes_2013} which used both VIMOS and FORS2 (ESO Large Programme 180.A-0776, PI: Almaini), and, as presented in \cite{maltby_identification_2016}, the targeted VIMOS observations of PSB candidates (ESO programme 094.A-0410, PI: Almaini). Following \cite{wilkinson_starburst_2021}, we limit our spectroscopic sample to galaxies with the most secure $z_{\mathrm{spec}}$ measurements. We also use these spectroscopic redshifts in place of our photometric redshifts, where available. We also remove a small number of X-ray emitting AGN identified by \cite{almaini_no_2025}, as the presence of AGN may impact our spectral line measurements leading to potentially unreliable spectroscopic classifications.

\section{Super-Colour Analysis}
\label{section:SC Analysis}
\subsection{Galaxy Classifications}
We follow the methodology of \cite{wild_new_2014} and \cite{wilkinson_starburst_2021} and apply a Principal Component Analysis (PCA) procedure to the spectral energy distributions (SEDs) of galaxies within the PRIMER-UDS field. This technique involves deriving a linear combination of three eigenspectra, with corresponding amplitudes referred to as `supercolours' (SCs), which when added to a mean SED can account for the majority of the variance in galaxy SEDs. These eigenvectors are derived from a library of tens of thousands of \cite{bruzual_stellar_2003} stellar synthesis models. Galaxies can then be classified as one of four populations using their position in the SC parameter space: star-forming, passive, dusty star-forming, or post-starburst galaxies. The supercolours do not directly correspond to physical properties, but SC1 is associated with the $R$-band weighted mean stellar age and dust content of a galaxy; SC2 is correlated with the fraction of stellar mass formed within the last Gyr and also helps break the dust-age degeneracy; and SC3 is used to break the degeneracy in SC2 between metallicity and the proportion of stellar mass formed in bursts in the last Gyr \citep{wild_new_2014}.

\begin{figure}
	\includegraphics[width=\columnwidth]{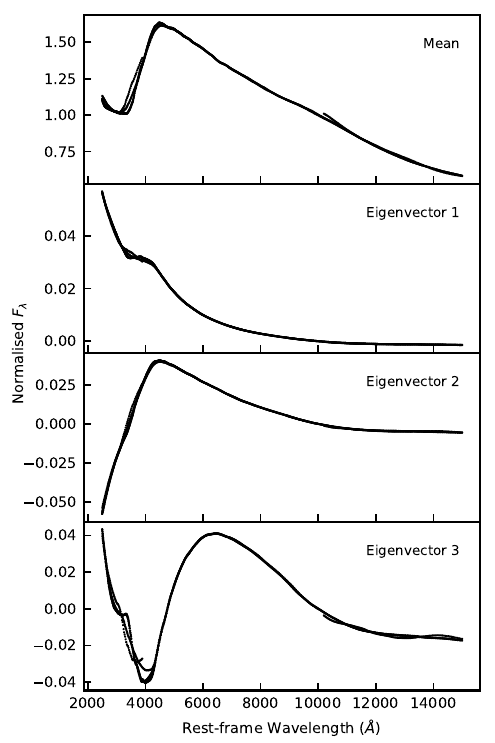}
    \caption{The eigenbasis generated in the PRIMER SC analysis (see Section~\ref{section:SC Analysis}). The eigenvectors are added or subtracted from the mean array in order to recreate the galaxy SEDs. The points represent the weight applied to a given filter for galaxies at different redshifts, and are placed at the central rest-frame wavelength of the filter. There are multiple curves as the same rest-frame wavelength can be probed by different filters depending on redshift.}
    \label{fig:Eigenbasis}
\end{figure}

In this work we adopt the extended rest-frame wavelength range of 2500--15\,000\,\AA\ used by \cite{wilkinson_starburst_2021} for the generation of the eigenbasis. Decreasing the lower wavelength limit from 3000\,\AA\ (as used in \citealt{wild_new_2014}) to 2500\,\AA\ allows the inclusion of our $F090W$ band measurements for galaxies at $z\sim2$, which improves the accuracy of their classifications due to the additional constraint at the blue end of their rest-frame SEDs. Fig.~\ref{fig:Eigenbasis} shows the eigenbasis used in this work, with the multiple curves reflecting that galaxies at different redshifts can have the same rest-frame wavelength observed by different filters, which have different transmission functions. We note that these eigenbases differ subtly from those used in both \cite{wild_new_2014} and \cite{wilkinson_starburst_2021}, due to the different filter sets used.

\begin{figure}
	\includegraphics[width=\columnwidth]{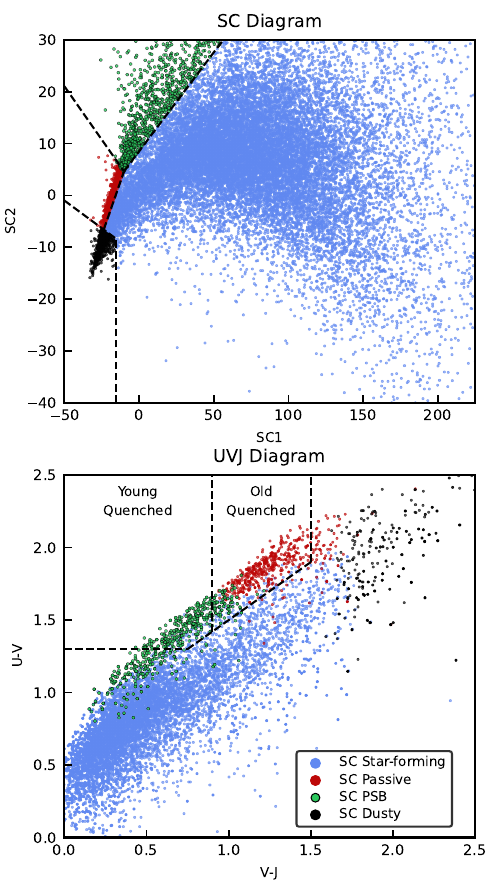}
    \caption{The top panel shows the PRIMER SC diagram for the galaxies in the PRIMER observations ($F200W$ < 26.5 mag), with the black dashed lines representing the boundaries used to classify galaxies into star-forming (blue), passive (red), PSBs (green) and dusty star-forming (black). The bottom panel illustrates the corresponding rest-frame UVJ diagram, with galaxies coloured by their \emph{JWST} SC classifications. The black dashed line shows the UVJ quiescent-selection criteria used by \protect\cite{whitaker_large_2012}.}
    \label{fig:JWST SC selection}
\end{figure}

Fig.~\ref{fig:JWST SC selection} illustrates how the galaxies are classified based on their position in the PRIMER SC diagram, and also shows how these SC classifications compare with the rest-frame UVJ classification criteria from \cite{whitaker_large_2012}. Over the redshift interval $0.5 < z < 3.0$, the SC method identifies 19\,587 star-forming galaxies, 473 dusty star-forming galaxies, 571 passive galaxies, and 807 PSBs. There are also 1750 galaxies that are designated `unclassified' due to extreme SC values that lie outside the classification region in SC-space. The majority of these unclassified galaxies are close to our magnitude limit, with large photometric uncertainties and correspondingly large SC errors.

We determined the SC classification boundaries by transforming those used in the UDS DR11 into the PRIMER SC space. We adopt the method used by \cite{wilkinson_starburst_2021}, who applied a similar transformation from the UDS to COSMOS filter sets. This was done by constructing mock SEDs along the UDS SC1--SC2 boundaries in step sizes of $\Delta$SC1= 0.1. The values of SC3 used to create the mock SEDs were chosen by fitting a second order polynomial through the SC3--SC1 plane of the SC values of real UDS DR11 galaxies that are within 0.5 SC2 units of each of the boundaries. The inclusion of an SC3 value when generating the mock SEDs only has a minimal effect on the transformation, as expected. These mock SEDs are then projected onto the PRIMER eigenvectors to derive the new classification boundaries by fitting linear lines of best fit. 

The SC classifications agree well with rest-frame UVJ classifications \citep{whitaker_large_2012}, with 99 per cent of the SC-selected star-forming and dusty star-forming galaxies lying outside the quenched region of the UVJ diagram, and 84 per cent of the passive galaxies located in the old quenched region. Only 39 per cent of the SC-selected PSBs satisfy the `young quenched' UVJ criteria, however, 52 per cent of our PSBs lie in the overall quenched region, and the remaining PSBs predominately lie just beyond the lower left boundary of the quenched region, which has been suggested to be where young, rapidly quenched galaxies reside \citep[e.g.][]{belli_mosfire_2019}. We estimate the likelihood of contamination within our PSB samples in Section~\ref{section:Contamination}.

\subsection{Stellar Masses}
Stellar masses and other relevant properties (e.g.\ age, star-formation rate, burst fraction) are determined during the SC method.  This uses a Bayesian analysis, described in more detail in \cite{wild_evolution_2016}, which accounts for the degeneracy between physical parameters. In brief, the library of \cite{bruzual_stellar_2003} stellar synthesis models are fit using the first three supercolours, which returns a probability density function (PDF) for each property. For the galaxies in our sample, the value of each property is taken as the median of the PDF, and the $1\mathrm{\sigma}$ uncertainties are determined from the 16th and 84th percentiles of the PDF. Stellar masses are calculated assuming a Chabrier initial mass function, and the resulting stellar mass uncertainties, due to random scatter, are typically $\sim 0.1\, \mathrm{dex}$. See \cite{wild_evolution_2016} and \cite{almaini_massive_2017} for detailed discussions of the systematic uncertainties in the stellar masses determined from the SC method.

\subsection{Spectroscopic Comparison}
\label{section:Spectroscopic Comparison}

\begin{figure}
	\includegraphics[width=\columnwidth]{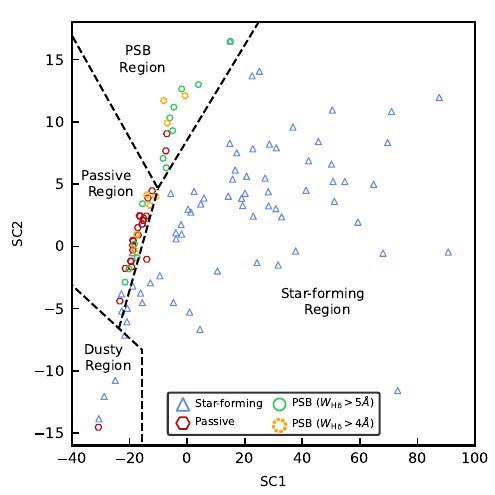}
    \caption{A PRIMER SC diagram showing the subset of galaxies which have spectroscopic classifications. Galaxies are displayed according to their spectroscopic classification: star-forming (blue triangles; $W_{\mathrm{[OII]}} < -10$\,\AA), passive (red hexagons; $W_{\mathrm{[OII]}} > -10$\,\AA, $W_{\mathrm{H\delta}} < 5$\,\AA), and PSBs using both the strict criteria (green circles; $W_{\mathrm{[OII]}} > -10$\,\AA, $W_{\mathrm{H\delta}} > 5$\,\AA) and the relaxed criteria (orange dashed circles; $W_{\mathrm{[OII]}} > -10$\,\AA, $W_{\mathrm{H\delta}} > 4$\,\AA).\ The black dashed lines show the boundaries used to classify galaxies photometrically using their SC values.}
    \label{fig:Spectral Classifications in SC-space}
\end{figure}

We present a comparison of the classifications from the SC analysis and the spectroscopic data described in Section~\ref{section:Spectroscopic Data}. The SC method has previously been robustly verified using extensive spectroscopy from across the wider UDS field \citep{maltby_identification_2016,wilkinson_starburst_2021}. Limited spectra are available in the PRIMER-UDS field. Therefore, the spectroscopic comparison here is provided for completeness and as a robustness test for consistency with \cite{maltby_identification_2016} and \cite{wilkinson_starburst_2021}. Fig.~\ref{fig:Spectral Classifications in SC-space} shows where galaxies with spectroscopic classifications lie in SC-space. We determine the spectral classifications of galaxies using the measured equivalent widths of $\mathrm{[O\ II]}$ ($W_{\mathrm{[OII]}}$) and $\mathrm{H\delta}$ ($W_{\mathrm{H\delta}}$). We apply the criterion of $W_{\mathrm{[OII]}} < -10$\,\AA\ to select galaxies that are star-forming, since the presence of $\mathrm{[O\ II]}$ emission lines are an indicator of ongoing star-formation and/or AGN activity \citep{yan_origin_2006}. We find that the SC and spectroscopic classifications agree extremely well with the separation of star-forming ($W_{\mathrm{[OII]}} < -10 $\,\AA) and quenched ($W_{\mathrm{[OII]}} > -10 $\,\AA) galaxies. We further divide the quenched spectroscopic sample into PSB and passive galaxies using $W_{\mathrm{H\delta}}$ (a marker of PSBs; \citealt{goto_catalogue_2007}). Similar to \cite{maltby_identification_2016} and \cite{wilkinson_starburst_2021}, we select PSBs using $W_{\mathrm{H\delta}} > 5$\,\AA, which gives PSB completeness and purity rates of $\sim62$$\pm13$ per cent (8/13) and $\sim57$$\pm13$ per cent (8/14), respectively, assuming binomial uncertainties. This completeness is comparable to those calculated for the UDS DR8 and DR11, however, we find a lower purity, which is likely due to low-number statistics. We note that there are multiple passive galaxies in the PSB region that have $4 < W_{\mathrm{H\delta}} < 5$\,\AA, which indicates that they have PSB-like features, but lie just below our formal classification criterion. Since the division between PSBs and the older passive population is somewhat arbitrary and the criteria used can vary in the literature \citep[e.g.][]{vergani_kgalaxies_2010, zhang_desi_2024, soto_catalog_2025}, we also apply a slightly lowered PSB criterion of $W_{\mathrm{H\delta}} > 4$ to identify galaxies with slightly less prominent PSB features. This adjusted PSB selection criteria results in a $52$$\pm11$ per cent (11/21) PSB completeness and $\sim79$$\pm11$ per cent (11/14) purity, which is consistent with the findings of \cite{maltby_identification_2016} and \cite{wilkinson_starburst_2021}. 

We conclude that the SC analysis provides excellent agreement with spectroscopic separation of star-forming and quenched galaxies, and reasonable agreement on the separation of PSBs within the quenched population. We note that the spectroscopic identification of PSBs is not necessarily superior to photometric selection, given the very distinctive broad-band SEDs of recently quenched galaxies (see \citealt{wild_star_2020}, for a discussion). We also compare the output of our SC analysis with that of the UDS DR11 \citep{wilkinson_starburst_2021} in Section~\ref{section:UDS Comparison} as a further test of our photometric classifications.
 
\subsection{Comparison with the UDS PCA analysis}
\label{section:UDS Comparison}
We compare the output of the PCA analysis from the PRIMER catalogue with the (largely) ground-based UDS classifications described in \cite{wilkinson_starburst_2021}. These catalogues are based on completely independent photometry using different instrumentation, with photometry from \textit{JWST} and {\em HST} used in this work, while \cite{wilkinson_starburst_2021} used ground-based plus \textit{Spitzer} photometry. Since 10 of the 12 bands used by \cite{wilkinson_starburst_2021} are from ground-based telescopes, hereafter we refer to this as a `ground-based' study, to distinguish from the entirely space-based photometry used in our work. Therefore, comparing the results of our PRIMER PCA output with that of the ground-based UDS for the overlapping galaxies provides a good test of robustness. For the purposes of this comparison, we exclude galaxies with $\frac{\Delta z}{1+z_{\rm PRIMER}} > 0.1$, in order to remove the effect of inconsistent redshifts between the UDS DR11 and our PRIMER catalogue. This removes $\sim15$ per cent of the overlapping galaxies, leaving a sample size of 6618 galaxies.

\begin{figure}
	\includegraphics[width=\columnwidth]{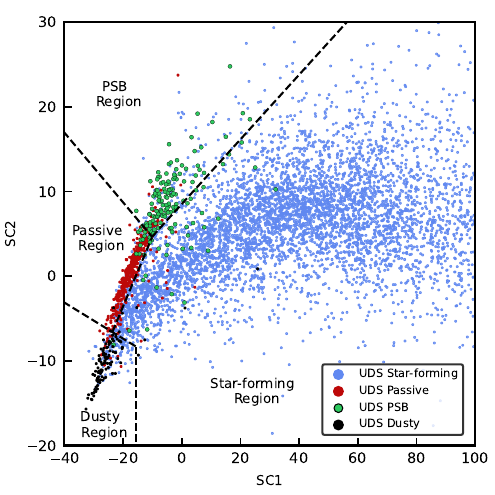}
    \caption{A PRIMER SC diagram showing the galaxies that are present in both the ground-based UDS DR11 and PRIMER catalogues, with similar redshifts (see Section~\ref{section:UDS Comparison} for details). The colours of the points represent the UDS DR11 SC classifications, with star-forming, passive, dusty and PSB populations represented by blue, red, black and green, respectively.}
    \label{fig:UDS galaxies in JWST SC space}
\end{figure}

Fig.~\ref{fig:UDS galaxies in JWST SC space} shows the agreement between our PRIMER classifications and the previous ground-based UDS classifications by plotting the overlapping galaxies in PRIMER SC-space, colour-coded by their UDS classifications. There is generally very good agreement, however, there are a number of galaxies whose UDS and PRIMER classifications differ, primarily by UDS galaxies moving out of the star-forming region. This movement of galaxies between SC classifications is not unexpected, as these scattered galaxies are generally close to the UDS DR11 magnitude limit, have large UDS SC errors, and a similar fraction of galaxies scatter out of each UDS SC region. This figure also shows that the boundaries have been translated correctly from the UDS SC-space, as most of the UDS-classified PSB (69 per cent) and passive (81 per cent) galaxies retain their classification in the PRIMER SC output.

\begin{figure}
	\includegraphics[width=\columnwidth]{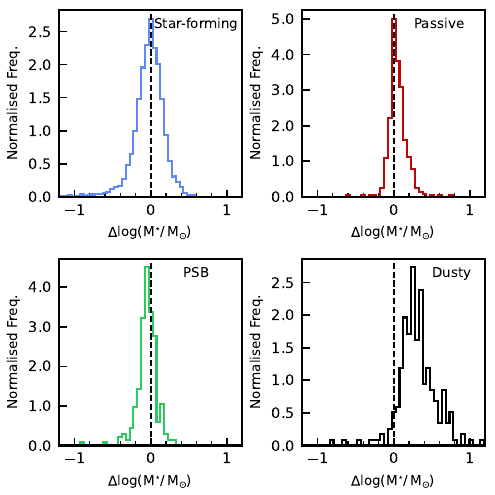}
    \caption{Histograms illustrating the difference in stellar masses between the PRIMER and ground-based UDS DR11 SC outputs for the overlapping galaxies between the two catalogs, split by their PRIMER SC classifications. Comparisons for star-forming (top left; blue), passive (top right; red), PSB (bottom left; green) and dusty star-forming galaxies (bottom right; black) are shown. Positive $\Delta \log(M_*\,/\,{\rm M_\odot})$ correspond to a galaxy's PRIMER stellar mass being higher than its stellar mass calculated from the UDS DR11. A vertical black dashed line at $\Delta \log(M_*\,/\,{\rm M_\odot}) = 0$ is included for reference.}
    \label{fig:JWST vs UDS masses Histogram}
\end{figure}

Fig.~\ref{fig:JWST vs UDS masses Histogram} shows how the stellar masses for the overlapping galaxies change between the UDS DR11 and PRIMER PCA analysis for each PRIMER sample. Aside from the dusty population, there is very good agreement between the masses from the UDS and PRIMER PCA analyses. However, we find that the masses of the dusty population are higher in our PRIMER analysis, which appears to be due to a combination of two key factors. Firstly, the majority of these galaxies have higher PRIMER photometric redshifts than those from the UDS DR11, which boosts their stellar masses accordingly. Secondly, further inspection reveals that many of these galaxies appear to be in relatively crowded regions, so the 2\arcsec\ diameter aperture used in the UDS ground-based analysis suffered from more contamination from nearby star-forming galaxies, leading to the lower assumed obscuration values and stellar masses. We do not expect these differences in stellar mass to have a significant impact on the stellar mass functions, as the dusty class make up only $\sim$ 2 per cent of the overall star-forming sample used in this work (see Section~\ref{section:Sample Selection} for details).

\subsection{Sample Selection}
\label{section:Sample Selection}
The primary samples used in this work are those classified using the PRIMER SC analysis, and for the purposes of this work we merge the dusty star-forming class with the typical star-forming population, to create a single overall star-forming sample.  This is to remain consistent with previous work using the SC analysis \citep[e.g.][]{wilkinson_starburst_2021, taylor_role_2023}. In addition, we also briefly analyse rest-frame UVJ-selected quenched and star-forming samples using the criteria from \cite{whitaker_large_2012}, to allow for a more direct comparison with previous studies that use these criteria.

\begin{table}
\centering
\begin{minipage}{70mm}
\centering
 \caption{Sample sizes for each SC population at each redshift.}
 \label{tab:sample sizes}
 \begin{tabular}{lccc}
  \hline
   & Star-forming & Passive & Post-starburst\\
  \hline
  $0.5 < z < 0.75$ & 2069 & 181 & 240\\
  $0.75 < z < 1.25$ & 4407 & 131 & 211\\
  $1.25 < z < 1.75$ & 4293 & 143 & 197\\
  $1.75 < z < 2.25$ & 4598 & 52 & 83\\
  $2.25 < z < 3.0$ & 4693 & 64 & 76\\
  \hline
 \end{tabular}
 \end{minipage}
\end{table}

We split the galaxies into 4 redshift bins between $z = 0.75$ and $z = 3$ to allow for a comparison with \cite{santini_stellar_2022} and \cite{hamadouche_jwst_2025}, and we also include an additional bin of $0.5 < z < 0.75$. The size of each population in each redshift bin is shown in Table~\ref{tab:sample sizes}. We calculate the 90 per cent stellar mass completeness limits for all of our samples by using the method of \cite{pozzetti_zcosmos_2010}. In brief, for a given sample, we calculate the stellar mass each galaxy would have if their apparent magnitude is shifted to our magnitude limit ($F200W = 26.5$), henceforth referred to as limiting masses. We then divide the galaxies into redshift bins of width 0.25 in the range $0.5 < z < 3.0$. We calculate the mass below which 90 per cent of the limiting masses lie for the 40 per cent faintest galaxies within each bin, before fitting second-order polynomials to obtain a functional form for the completeness as a function of redshift. We use the 40 per cent faintest galaxies in this calculation instead of 20 per cent used in \cite{pozzetti_zcosmos_2010} to ensure that there are sufficient numbers of galaxies in each redshift bin, particularly at higher redshift. This modification has a minimal impact on the resulting 90 per cent mass completeness curves. The calculation is limited to the faintest galaxies within each redshift bin in order to obtain a more realistic estimate for the mass-to-light ratio of galaxies close to the mass completeness limit.

\begin{figure}
	\includegraphics[width=\columnwidth]{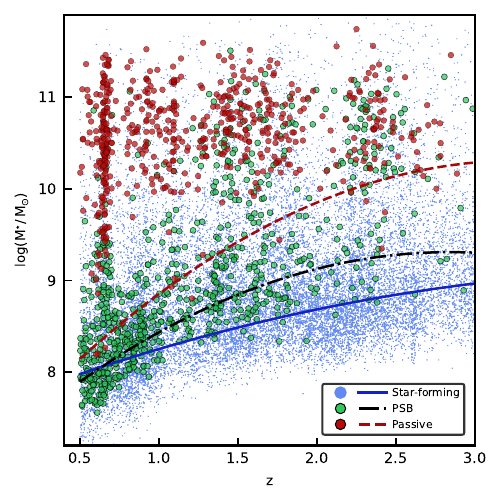}
    \caption{Stellar mass vs. redshift for the galaxies from the PRIMER SC analysis, with the star-forming, passive and PSB galaxies denoted in blue, red and green, respectively. The solid blue, red dashed and black dashed-dotted lines show the 90 per cent mass completeness curves for the star-forming, passive and PSB populations, respectively.}
    \label{fig:Mass Completeness}
\end{figure}

Fig.~\ref{fig:Mass Completeness} shows the mass distribution against redshift for each of the three SC-selected samples used in this work, along with their 90 per cent mass completeness curves.

\section{Stellar Mass Functions}

In this section, we present the galaxy stellar mass functions for our samples, covering a redshift range of $0.5 < z < 3.0$. This allows us to investigate the prevalence of each galaxy population in various mass regimes across cosmic time, and to explore the evidence for a low-mass quenching pathway out to cosmic noon.

\subsection{SC-Selected Samples}
\begin{figure*}
	\includegraphics[width=17.5cm]{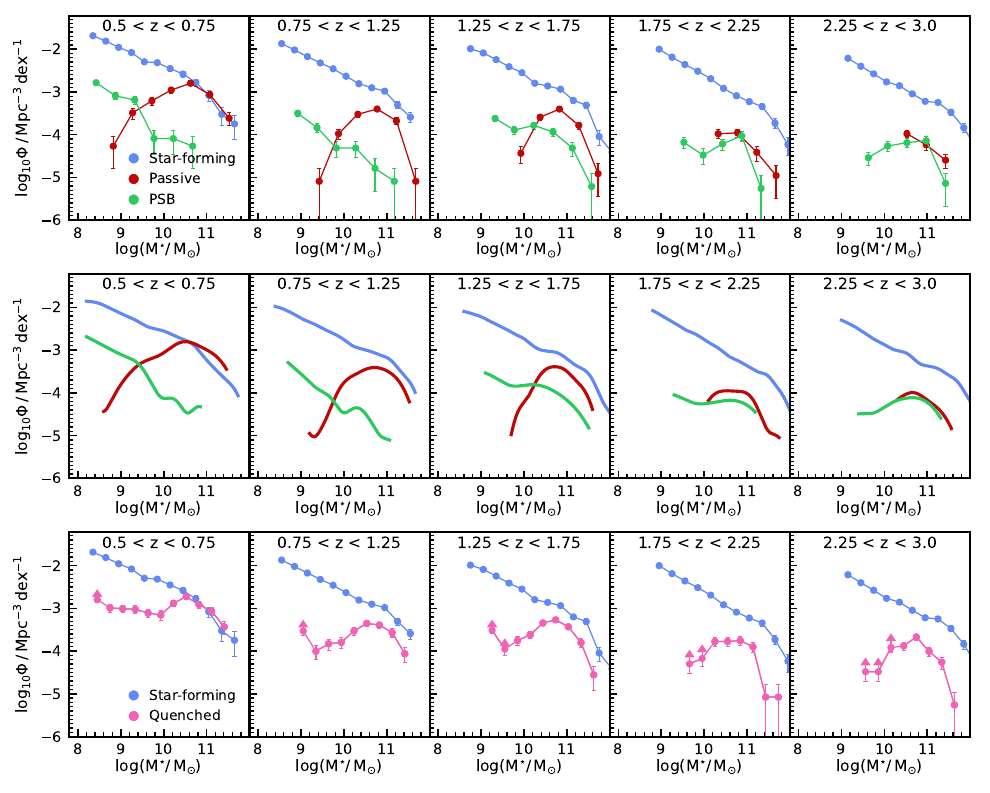}
    \caption{ \emph{Top row}: global stellar mass functions for the star-forming, passive, and PSB populations from the PRIMER SC analysis. \emph{Middle row}: alternative smoothed mass functions based on applying a kernel density estimation (KDE) analysis.  \emph{Bottom row}: global stellar mass functions for star-forming galaxies, and the combined quenched population of passive and PSB galaxies. The mass functions and KDEs of the individual classes are shown above the 90 per cent mass completeness limit, evaluated at the upper limit of each redshift bin for each population. The combined passive population is shown above the PSB 90 per cent mass completeness limit, with the bins that fall below the passive 90 per cent mass completeness limit denoted by lower limits. This shows that the low-mass quenched galaxy population is dominated by PSBs out to at least $z \sim 1.5$}
    \label{fig:SC Mass Functions}
\end{figure*}

In Fig.~\ref{fig:SC Mass Functions}, we present the stellar mass functions for the SC-selected star-forming, passive and PSB populations as a function of redshift. We also present the mass functions for the overall quenched population by combining the PSB and passive mass functions. In addition to the standard binned mass functions, we also include the results of applying a kernel density estimation (KDE; \citealt{rosenblatt_remarks_1956}; \citealt{parzen_estimation_1962}), to ensure that any conclusions drawn are not sensitive to binning. The KDE is determined by representing each galaxy with a Gaussian kernel, which are then summed together in order to produce a probability density function. The kernel bandwidth is selected using Scott’s rule \citep{Scott1992}, for which the bandwidth factor is given by
\begin{equation}
h = n^{-1/(d+4)},
\end{equation}
where $n$ is the number of galaxies and $d$ is the dimensionality of the data; in one dimension this corresponds to a Gaussian kernel with standard deviation $\sigma = h\,\sigma_{\log M_{\ast}}$. These are then normalised to the units of the mass function. In order to minimise the effect of the boundary bias problem (e.g. \citealt{muller_multivariate_1999}), which leads to an underestimation at the boundaries, we generate the KDE using galaxies at all masses but truncate it at the 90 per cent mass completeness limit.

Inspection of the mass functions in Fig.~\ref{fig:SC Mass Functions} reveals the expected trends for the star-forming and quenched mass functions \citep[e.g.][]{muzzin_evolution_2013}: the quenched mass function shows a build-up in number density over cosmic time, particularly at the high-mass end, and the shape of the star-forming mass function remains largely consistent at all redshifts. Work prior to \textit{JWST} had detected a low-mass upturn in the PSB mass function below $z = 1.5$ \citep[e.g.][]{wild_evolution_2016, taylor_role_2023}, but an upturn was unable to be detected at higher redshifts due to mass incompleteness. We detect this upturn\footnote{The term `upturn' refers to an excess of low-mass galaxies compared to that which would be expected from a single Schecter function with a high characteristic mass.} in the PSB mass function below $z = 1.5$, but we also see evidence of an upturn to $z = 2.25$. A similar upturn has been reported for the overall quenched mass function by  \cite{santini_stellar_2022}, \cite{hamadouche_jwst_2025} and \cite{shuntov_stellar_2026}. This distinct upturn below $M_{\ast}=10^{10}\,\mathrm{M_{\odot}}$ is indicative of an alternative quenching pathway for lower-mass galaxies. We likewise see evidence for this low-mass upturn in the overall quenched population out to $z=2.25$, but interestingly we do not see this trend in the older passive population (i.e. without PSBs) at any redshift. We conclude that recently and rapidly quenched galaxies are the primary contributor to the low-mass upturn seen in the overall quenched population. Furthermore, given that PSBs are typically expected to evolve into passive galaxies without any significant increase in stellar mass, the lack of low-mass passive galaxies at lower redshifts, despite the presence of low-mass PSBs at higher redshift, provides further evidence that high- and low-mass PSBs follow different evolutionary pathways. We investigate this further in Section~\ref{section:Building up passive population}.

We note that the low-mass upturn in the mass functions of photometrically selected quenched galaxies could (in principle) be affected by the misclassification of star-forming galaxies, which we investigate in Section~\ref{section:Contamination}. We find that the impact on our primary conclusions is minimal.

\subsection{UVJ-selected Samples}
\begin{figure*}
	\includegraphics[width=17.5cm]{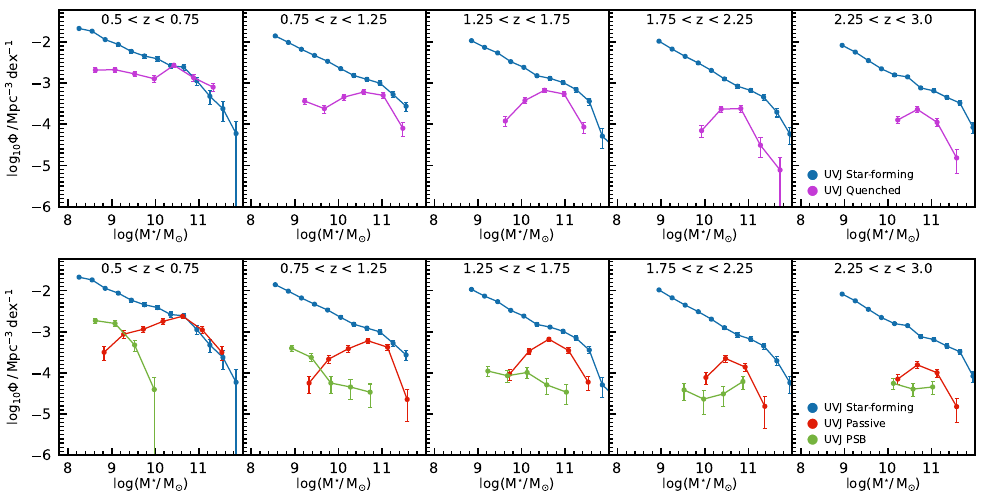}
    \caption{\emph{Top row}: global stellar mass functions for UVJ-selected star-forming (blue) and quenched (purple) populations from the PRIMER-UDS field using the criteria from \protect\cite{whitaker_large_2012}. \emph{Bottom row}: global stellar mass functions for UVJ-selected star-forming (blue), old quenched (red), and young quenched (green) populations from the PRIMER SC analysis using the criteria from \protect\cite{whitaker_large_2012}. The mass functions are shown above the 90 per cent mass completeness limit for each population, evaluated at the upper limit of each redshift bin.}
    \label{fig:UVJ Mass Functions}
\end{figure*}

In Fig.~\ref{fig:UVJ Mass Functions}, we show the mass functions for our galaxy sample classified using the rest-frame UVJ boundaries from \cite{whitaker_large_2012} This enables a more direct comparison with studies that do not use PCA classification. The `young quiescent' population from \cite{whitaker_large_2012} is analogous to the SC-selected PSB sample (see \citealt{wild_new_2014} and \citealt{almaini_massive_2017}, for comparisons). We find that our UVJ-selected mass functions follow the same general trends as those using SC-selected samples. In particular, we find a low-mass upturn at, or just below, $M_{\ast}=10^{10}\,\mathrm{M_{\odot}}$ in the `young quenched' population out as far as $z = 2.25$. We do not detect any upturn in the overall passive UVJ-selected population in the $1.75 < z < 2.25$ redshift bin, in contrast to the findings of \cite{hamadouche_jwst_2025}. However, our mass functions do not extend as low in stellar mass due to our conservative magnitude limit. \cite{hamadouche_jwst_2025} detect an upturn below $M_{\ast}=10^{9.5}\,\mathrm{M_{\odot}}$ in the redshift range $1.75 < z < 2.25$, below the completeness limit of our current study. Therefore, we conclude that our results are not at odds with those of \cite{hamadouche_jwst_2025}. Overall, our UVJ analysis supports the conclusions of the SC analysis and underscores the importance of distinguishing recently quenched galaxies from the broader quenched population, as we find no evidence of an upturn in the older quenched population at any redshift. We conclude that the low-mass upturn in the quenched galaxies mass functions is largely driven by the addition of rapidly quenched galaxies.

\subsection{Potential Sample Contamination}
\label{section:Contamination}
\begin{figure*}
	\includegraphics[width=17.5cm]{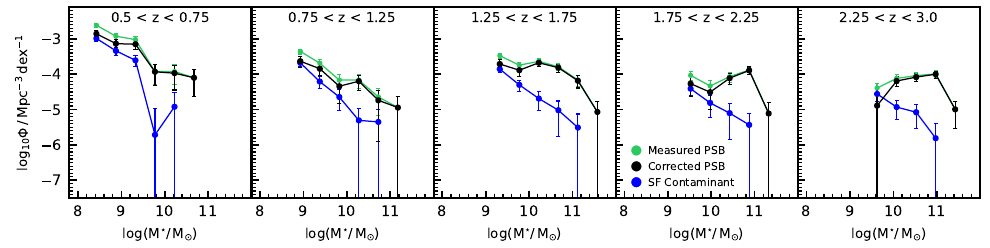}
    \caption{The effect on the PSB mass function of removing potential contamination from star-forming galaxies. The original PSB mass functions are shown in green. An estimate of the contamination from star-forming galaxies in the PSB population, calculated from resampling SC values within errors from 10\,000 iterations, is shown in blue. The result of subtracting the contaminant mass functions from our original PSB mass functions are shown in black. The mass functions are shown above the 90 per cent mass completeness limit of the original PSB population, evaluated at the upper limit of each redshift bin.}
    \label{fig:PSB Contaminant Mass Functions}
\end{figure*}

Classifying galaxies photometrically using either rest-frame UVJ or SC selection criteria involves applying discrete boundaries to colours with intrinsic uncertainties. Therefore, galaxies lying close to the classification boundaries will have a non-negligible probability of being incorrectly classified. As star-forming galaxies are significantly more numerous than passive and post-starburst galaxies, even a small fraction of star-forming galaxies being misclassified can skew the results of the other samples. We would expect contamination to be more prevalent at lower masses, at a given redshift, as fainter galaxies typically have larger photometric errors. Therefore the low-mass upturn in our PSB mass functions could (in principle) be affected by contamination from star-forming galaxies. To investigate the potential impact of this contamination, we perform a simple test by resampling the SC values of the star-forming population, assuming a Gaussian distribution using the uncertainties in their SC values. We then calculate the mass function of the galaxies that move into the PSB classification region, which we refer to as `contaminant' galaxies. In Fig.~\ref{fig:PSB Contaminant Mass Functions}, we compare our original PSB mass functions with the average contaminant mass function from 10\,000 iterations, in addition to showing the `corrected' PSB mass functions, calculated by subtracting the average contaminant mass functions from our original PSB mass functions. As expected, we find that the shape of the contaminant mass functions match those of the original star-forming samples. However, the estimated contaminants do not contribute significantly to the PSB mass functions, and do not appear to be the cause of the low-mass upturn. Notably, the `corrected' PSB mass functions still show evidence of an upturn below $z < 2$. However, the slight decrease in the upturn after accounting for the effect of star-forming contaminants could cause a minor improvement in the overprediction of low-mass passive galaxies discussed in Section~\ref{section:Modeling PSB contribution}. Based on these mass functions, we estimate that approximately $\sim 35$ per cent of our low-mass PSB samples ($M_{\ast}<10^{10}\,\mathrm{M_{\odot}}$) are potentially contaminant galaxies (above the 90 per cent mass completeness limits), with a negligible contribution at high stellar mass.

\section{Building up the passive galaxy mass function}
\label{section:Building up passive population}
We now explore the growth of the passive galaxy mass function across cosmic time by applying a simple model which assumes that all PSBs evolve into passive galaxies. This allows us to probe for differences in the evolution of the quenched galaxy populations in various mass regimes since cosmic noon.

\subsection{Modelling the contribution from post-starburst galaxies}
\label{section:Modeling PSB contribution}

The PSB phase is expected to be visible for up to $\sim $1 Gyr in our SC-space, given the timescale over which A/F-type stars will dominate the galaxy SED. This is similar to the cosmic time spanning each redshift bin. We predict the passive mass function in a given redshift bin by using a combination of the passive and PSB mass functions in the higher adjacent redshift bin:
\begin{equation}
\label{eqn:Mass function prediction}
 \phi_{\mathrm{pred}}(M_{*})= \phi_{\mathrm{PAS,\ hi-z}}(M_{*}) + \frac{\Delta t_{\mathrm{cos}}}{t_{\mathrm{PSB}}} \times \phi_{\mathrm{PSB,\ hi-z}}(M_{*})\ ,
\end{equation}
\noindent where $\phi_{\mathrm{pred}}(M_{*})$ is the predicted passive mass function, and $\phi_{\mathrm{PAS,\ hi-z}}(M_{*})$ and $\phi_{\mathrm{PSB,\ hi-z}}(M_{*})$ are the passive and PSB mass functions in the higher redshift bin, respectively. The term $\Delta t_{\mathrm{cos}}$ is the cosmic time interval spanning the higher redshift bin, and $t_{\mathrm{PSB}}$ is the PSB visibility timescale. Following the analysis of \cite{wild_evolution_2016}, these predictions allow us to estimate the change in the passive mass function, if all the growth is due to the transition of PSBs onto the red sequence, ignoring the effect of mergers or rejuvenation.

\begin{figure*}
	\includegraphics[width=17.5cm]{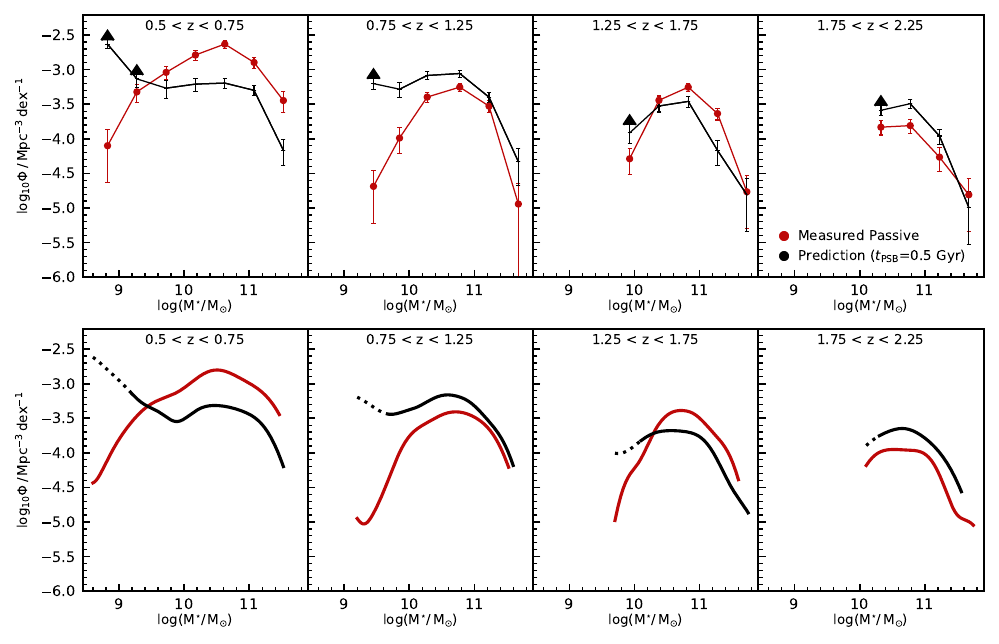}
    \caption{The observed passive galaxy mass functions (red) compared to the predicted mass functions (black), obtained by combining the passive and (scaled) PSB mass functions from the earlier redshift bin. The scaling of the PSB mass function accounts for the visibility time (assumed to be 500 Myr) and the cosmic time interval corresponding to that bin (see Section~\ref{section:Modeling PSB contribution}, for further details). The upper and lower panels show the binned and KDE mass functions, respectively.  All mass functions are shown down to the 90 per cent mass completeness limit of the measured passive mass function, with lower limits and dotted lines indicating where the predicted mass functions and KDEs are incomplete, respectively. We find that this simple model overpredicts the number density of low-mass passive galaxies ($M_{\ast} \lesssim 10^{10}\,\mathrm{M_{\odot}}$) at redshifts below $z\sim1.75$, which may indicate a different evolutionary pathway for high- and low-mass PSBs}
    \label{fig:Passive Mass Function Buildup}
\end{figure*}

In Fig.~\ref{fig:Passive Mass Function Buildup}, we show the comparison between the measured passive mass functions and our predicted passive mass functions, assuming a PSB visibility time of 500 Myr \citep{wild_evolution_2016}. We find that the shape of the predicted passive mass function largely matches that of the measured passive mass function at $M_{\ast} \gtrsim 10^{10}\rm\,M_{\odot}$ in all redshift bins, which is consistent with the findings of \cite{wild_evolution_2016} in the redshift range $0.5 < z < 1.5$. However, we find that we overpredict the number of passive galaxies at lower stellar masses at $z < 1.75$. At higher redshifts, we are unable to probe low enough stellar masses to determine if a similar overprediction occurs. This result is consistent with \cite{taylor_role_2023} who found that the quiescent number density growth rate is consistent with lower-mass PSBs [$9.7 < \log(M_\ast/\mathrm{M_\odot}) < 9.9$] having longer visibility times than higher mass PSBs [$\log(M_\ast/\mathrm{M_\odot}) > 10.2$] . This may suggest the assumption where post-starburst galaxies transition to the red-sequence uniformly in mass is incorrect, and that there may be at least one mass-dependent mechanism that reduces the rate at which low-mass PSBs join the passive population. We explore some possible explanations in the following sections.

\subsection{Mass-dependent quenching timescales}

\begin{figure*}
	\includegraphics[width=17.5cm]{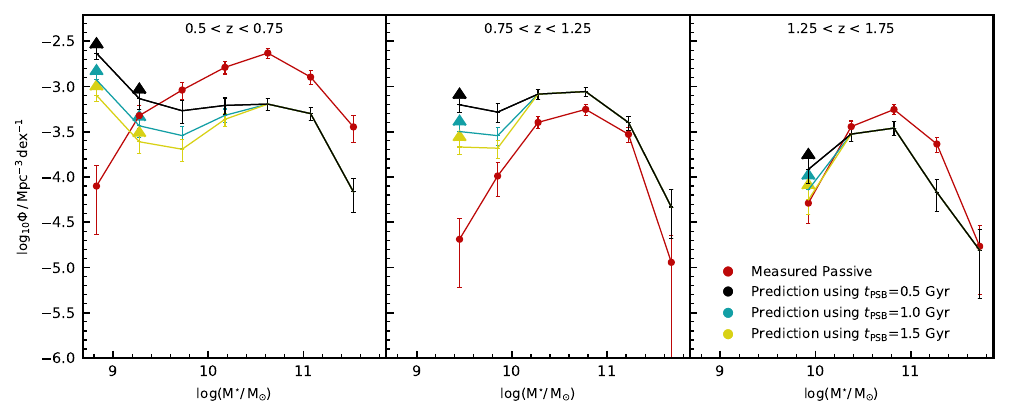}
    \caption{The measured and predicted passive galaxy mass functions, derived using the formalism described in Section~\ref{section:Modeling PSB contribution}, assuming a range of visibility times for low-mass PSBs ($M_{\ast} \lesssim 10^{10}\,\mathrm{M_{\odot}}$). We exclude the $1.75 < z < 2.25$ redshift bin as there are insufficient galaxies at $M_{\ast} < 10^{10}\,\mathrm{M_{\odot}}$ to explore the potential upturn. Increasing the visibility time of the PSB phase at low masses does not fully reconcile the overprediction, which suggests the presence of additional or alternative mechanisms that suppress the transition of low-mass PSBs to the passive population.}
    \label{fig:Mass Dependent Visibilities}
\end{figure*}

Our predictions assume that all PSBs are visible for a uniform length of time, regardless of mass. However, it has been found that high-mass PSBs ($M_{\ast}$ > $10^{10}\,\mathrm{M_{\odot}}$) have structures similar to those of high-mass passive galaxies but are much more compact, which suggests they experienced a violent quenching event, e.g.\ a gas-rich merger or dissipative disk collapse \citep{almaini_massive_2017, maltby_structure_2018}. In contrast, \cite{maltby_structure_2018} and \cite{maltby_multiwavelength_2026} found that low-mass PSBs ($M_{\ast}$ < $10^{10}\,\mathrm{M_{\odot}}$) have much lower S\'ersic indices than high-mass PSBs and are disc-dominated, which suggests that they underwent a more gentle quenching process, likely an environmental process such as gas stripping, minor mergers or harrassment. Therefore, if high- and low-mass PSBs undergo different quenching processes, then it is plausible that they quench over different timescales, which could affect the timescales over which the PSB phase is visible. \cite{wild_star_2020} performed detailed fitting to the optical spectra of high-mass PSBs ($M_{\ast} \gtrsim 10^{10}\,\mathrm{M_{\odot}}$) at $z \sim 1$ and found that the SC-measured post-starburst features are visible for $0.5$--$1$ Gyr in this population. However, \cite{skarbinski_jwst_2026}, using similar methods, finds tentative evidence that the SC PSB visibility timescales increase toward lower masses for galaxies with $M_{\ast} > 10^{10}\,\mathrm{M_{\odot}}$. They find a median visibility time of $\sim 340$ Myr for massive ($M_{\ast} > 10^{10.7}\,\mathrm{M_{\odot}}$) PSBs, and a median of $\sim 660$ Myr for lower mass ($M_{\ast} < 10^{10.7}\,\mathrm{M_{\odot}}$) PSBs. Therefore, it is plausible that low-mass ($M_{\ast} < 10^{10}\,\mathrm{M_{\odot}}$) PSBs may have even longer visibility times, should this trend extend below the mass regime investigated by \cite{skarbinski_jwst_2026}, which could be long as $\sim 1.5$ Gyr due to the lifetime of A/F-type stellar populations. In Fig.~\ref{fig:Mass Dependent Visibilities}, we explore the effect on our predictions of increasing the PSB visibility timescales for low-mass PSBs ($M_{\ast} \lesssim 10^{10}\,\mathrm{M_{\odot}}$). Unsurprisingly, we find that increasing the PSB visibility timescale at low masses significantly reduces our overprediction. In both redshift bins, the 1.5 Gyr visibility timescales most closely match the shape of the passive mass functions, which may provide further evidence for different quenching pathways for high- and low-mass PSBs. However, this does not fully correct the overpredictions. Given that 1.5 Gyr is at the upper end of theoretically possible PSB visibility timescales due to the lifetime of A/F-type stars, the typical value is likely to be below 1.0 Gyr \citep[e.g.][]{wild_star_2020, skarbinski_jwst_2026}. This is likely an upper limit on the effect of mass-dependent visibility timescales. Therefore, we conclude that there are likely additional mechanisms which preferentially affect low-mass galaxies to prevent a significant build-up in the number density of the low-mass passive population.

\subsection{Cluster Correction}

\begin{table}
\centering
\begin{minipage}{80mm}
\centering
\caption{The ratio of the predicted number density to the observed number density for high- ($M_{\ast} > 10^{10}\,\mathrm{M_{\odot}}$) and low-mass ($M_{\ast} < 10^{10}\,\mathrm{M_{\odot}}$) passive galaxies in each redshift bin, before and after the cluster corrections are applied.}
\label{tab:cluster correction}
\begin{tabular}{lcccc}
 \hline
  Epoch       & \multicolumn{2}{c}{Low-mass} & \multicolumn{2}{c}{High-mass}\\
  {}          & Uncorr. & Corr. & Uncorr. & Corr.\\
  \hline
  $0.5 < z < 0.75$  & $2.4$  & $10.7$ & $0.3$& $1.0$\\
  $0.75 < z < 1.25$ & $11.5$ & $2.3$  & $1.7$ & $0.6$\\
  $1.25 < z < 1.75$ & $2.0$  & $3.7$  & $0.7$ & $0.7$\\
  \hline
 \end{tabular}
\end{minipage}
\end{table}

Two large-scale galaxy structures lie within the PRIMER UDS field, one at $z = 0.65$ \citep{galametz_growing_2018} and another at $z = 1.62$ \citep{papovich_spitzer-selected_2010}, which may impact the results of our simple evolutionary models. We therefore repeat our predictions for the passive mass functions, assuming a uniform PSB visibility time of 500 Myr, with thin redshift slices removed at $z = 0.65$ and $z = 1.62$. We remove galaxies within $\frac{\Delta z}{(1+z)} = 0.057$, corresponding to the $\pm\,3 \sigma$ uncertainty in the photometric redshifts. 
In Table~\ref{tab:cluster correction} we show the ratio of the predicted number density to the observed number density for high- ($M_{\ast} > 10^{10}\,\mathrm{M_{\odot}}$) and low-mass ($M_{\ast} < 10^{10}\,\mathrm{M_{\odot}}$) passive galaxies in each redshift bin. We find that the magnitude of the overprediction of low-mass passive galaxies increases for the redshift bins containing the clusters after applying the cluster correction, whilst it decreases in the $0.75 < z < 1.25$ redshift bin. Therefore, although the presence of these large galaxy overdensities do impact the evolutionary trends, the presence of a significant overprediction of low-mass passive galaxies remains.

\subsection{The role of mergers, rejuvenation and stochastic star-formation}
\label{section:Implications}

Performing a further investigation into the cause of overprediction of low-mass passive galaxies is beyond the scope of this work. Instead, we discuss the potential nature of additional processes which could inhibit the build up of the low-mass passive population and present possible differences in the evolutionary pathways of high- and low-mass PSBs which may reconcile the apparent overproduction of the low-mass passive mass function.

Assuming that our overprediction is only due to mechanisms that impact the evolution of PSBs, we can use our mass functions to estimate the relative prevalence of these additional processes in the high- ($M_{\ast} > 10^{10}\,\mathrm{M_{\odot}}$) and low-mass ($M_{\ast} < 10^{10}\,\mathrm{M_{\odot}}$) PSB populations. We find that in the $0.5 < z < 0.75$ redshift bin these processes preferentially impact low-mass PSBs by $\sim 25$ times the rate of high-mass PSBs, assuming a single mass-independent PSB visibility time, or $\sim 13$ times if the typical visibility period of low-mass PSBs is double that of high-mass PSBs. In the $0.75 < z < 1.25$ redshift bin we find that these values are $\sim 3.1$ and $\sim 1.6$ times, respectively. These values in the $1.25< z < 1.75$ redshift bin are $\sim 4.7$ and $\sim 2.3$ times, respectively. We note that these estimates are also likely to be lower limits due to the incompleteness of our predictions at the lowest masses analysed at each redshift, indicated in Fig.~\ref{fig:Passive Mass Function Buildup} and Fig.~\ref{fig:Mass Dependent Visibilities}.

The lack of a significant build-up in the low-mass passive population could be accounted for through the rejuvenation of star-formation in low-mass quenched galaxies, and/or through mechanisms resulting in the removal of low-mass quenched galaxies, such as mergers and disruption. \cite{edward_stellar_2024} finds a relative overprediction of low-mass passive galaxies at $z \sim 1.3$ using the SMF of protoclusters at $2 < z < 2.5$, which they attribute to significant merging and/or disruption. In addition, \cite{harrold_role_2026} explored the effect of mergers, rejuvenation and PSB visibility times using a mock UDS lightcone generated from the {\sc L-galaxies} Semi Analytic Model. They found that low-mass ($M_{\ast} < 10^{10.5}\,\mathrm{M_{\odot}}$) PSB and passive galaxies in the redshift range $0.76 < z < 3.0$ are much more likely to merge after observation than those at high mass ($M_{\ast} > 10^{10.5}\,\mathrm{M_{\odot}}$). They also find that taking into account the effect of both rejuvenation and mergers reduces the contribution of PSBs to the growth of the passive galaxy mass function by a factor of 2. This suggests that both mergers and rejuvenation may play an important role in reducing the build up of the low-mass passive population.

Locally, there is evidence to suggest that a significant fraction of PSBs rejuvenate and return to the star-forming population \citep{dressler_imacs_2013, rowlands_evolution_2015, pawlik_origins_2018}. Furthermore, it has been suggested that tidal effects in clusters can trigger starbursts \citep{gendron_numerical_2025}, therefore low-mass PSBs may rejuvenate more frequently than those at high-mass, given that they typically reside in higher density environments \cite[e.g.][]{socolovsky_enhancement_2018, wilkinson_starburst_2021, taylor_role_2023}. In addition, it has been suggested that low-mass galaxies undergo stochastic star-formation, both locally \cite[e.g.][]{kauffmann_quantitative_2014}, and at higher redshifts \cite[e.g.][]{atek_star_2022, mintz_taking_2026}. This has also been observed in simulations \citep{shen_baryon_2014, flores_velazquez_time-scales_2021}. Therefore, it is possible that these galaxies may be identified as PSBs in our analysis while they are in between bursts of star-formation, as numerous low-mass galaxies believed to have bursty star-formation histories have been been found to have strong Balmer breaks \citep[a PSB signature; ][]{mintz_taking_2026}. This could mean that a significant fraction of our low-mass PSBs rejoin the star-forming population instead of transitioning permanently to the passive population.

Therefore, we conclude that high- and low-mass PSBs follow different evolutionary pathways, which requires further investigation. We propose a potential scenario where low-mass PSBs are quenched through slower, more gentle quenching mechanisms (possibly due to their environment) but are either prevented from transitioning to the older passive population through undergoing subsequent periods of rejuvenation and quiescence, or removed entirely through mergers or disruption.

\section{Summary and Conclusions}

We have investigated the mass functions of star-forming, passive, and PSB galaxies in the PRIMER-UDS field out to $z = 3$. These galaxies were selected through a photometric PCA technique providing a sample of $\sim850$ PSBs in the redshift range $0.5 < z < 3.0$. We compared the passive galaxy mass functions to the predicted passive mass functions, assuming all the growth is due to the addition of galaxies that pass through a PSB phase. Our key findings are summarised as follows:

\begin{enumerate}
    \item We confirm the presence of the recently discovered low-mass upturn in the overall quenched (i.e. passive + PSB) population out to $z \sim 2$ (e.g. \citealt{santini_stellar_2022}; \citealt{hamadouche_jwst_2025}; \citealt{shuntov_stellar_2026}). Our new results suggest that this upturn can be primarily attributed to recently quenched galaxies, as we find minimal evidence of an upturn in the older passive population (i.e., with PSBs removed).
    \vspace{0.1cm}
    \item We find an upturn at the low-mass end of the PSB mass function out to $z = 2.25$, which shows that a low-mass quenching route is active out to cosmic noon, potentially driven by environmental effects.
    \vspace{0.1cm}
    \item If we assume that all the growth in the passive mass function comes from PSBs with a uniform visibility time, we find a relative overprediction of galaxies at the low-mass end of the passive mass function. This suggests that there are factors that disproportionately affect the calculated rate at which low-mass PSBs transition into the passive population. Employing a longer visibility timescale for PSBs at $M_{\ast} \lesssim 10^{10}\,\mathrm{M_{\odot}}$ significantly reduces the overprediction of low-mass passive galaxies, which may indicate further evidence that high- and low-mass PSBs follow different quenching pathways. Employing mass-dependent visibility timescales does not fully correct the overprediction, which suggests the presence of additional processes which significantly inhibit the transition of low-mass PSBs into the passive population.
\end{enumerate}

Our work provides further evidence for a mass-independent quenching mechanism(s) that is active out to $z \sim 2$, possibly driven by environment, supporting the conclusions of \cite{santini_stellar_2022}, \cite{hamadouche_jwst_2025} and \cite{shuntov_stellar_2026}. \cite{santini_stellar_2022} compare their observational results with a range of hydrodynamical simulations and Semi-Analytic Models (SAMs). In brief, despite differences in the precise galaxy mass functions, the models generally showed evidence for a low-mass upturn in the passive galaxy mass function out to at least $z \sim 2$. For the models where they extracted environmental information, they found that this low-mass upturn was primarily due to satellite galaxies. Observationally, further evidence for environmental quenching at cosmic noon has been found from studies of galaxy clustering. These reveal that low-mass PSBs reside in very massive dark matter halos to at least $z\sim 1.5$ \citep{wilkinson_starburst_2021}, and the evidence for a low-mass excess of passive galaxies in protoclusters compared to the field at $z\sim2$ \citep{edward_stellar_2024}.

The key new finding in our work is the discovery that recently quenched galaxies account for the bulk of the low-mass upturn in the passive galaxy mass function to $z\sim2$. This adds to the body of evidence suggesting that high-and low-mass galaxies typically follow different quenching pathways. Further work investigating the differences between the high- and low-mass PSBs, such as in morphology and environment, may shed further light on which quenching mechanisms dominate in each mass regime.

\section*{Acknowledgements}

TdL acknowledges the support of STFC studentship ST/X508639/1. OA acknowledges the support of STFC grant ST/X006581/1. VW acknowledges the Science and Technologies Facilities Council (ST/Y00275X/1) and Leverhulme Research Fellowship (RF-2024-589/4). MS and KR gratefully acknowledge support from the NASA Astrophysics Data Analysis Program (ADAP) under grant 80NSSC23K0495. ET and ACC acknowledge support from a UKRI Frontier Research Guarantee Grant (PI Carnall; grant reference EP/Y037065/1). JSD and DJM acknowledge the support of the Royal Society through the award of a Royal Society Professorship to JSD. We extend our gratitude to the staff at United Kingdom Infrared Telescope for their tireless efforts in ensuring the success of the UDS project. We also wish to recognize and acknowledge the very significant cultural role and reverence that the summit of Mauna Kea has within the indigenous Hawaiian community. We were most fortunate to have the opportunity to conduct observations from this mountain. For the purpose of open access, the authors have applied a creative commons attribution (CC BY) to any journal-accepted manuscript. This work is based in part on observations from ESO telescopes at the Paranal Observatory (programmes 180.A-0776, 094.A-0410 and 194.A-2003).

\section*{Data Availability}

The data forming the basis of this work is available from public archives, including the Mikulski Archive for Space Telescopes (https://mast.stsci.edu/), and the UKIDSS UDS web page (https://www.nottingham.ac.uk/astronomy/UDS/). A public release of the processed data, galaxy properties, and photometric redshifts is in preparation. Data can also be obtained on reasonable request to the corresponding author.



\bibliographystyle{mnras}
\bibliography{TdL_bibtex} 

@ARTICLE{Taylor2024,
       author = {{Taylor}, Elizabeth and {Maltby}, David and {Almaini}, Omar and {Merrifield}, Michael and {Wild}, Vivienne and {Rowlands}, Kate and {Harrold}, Jimi},
        title = "{High-velocity outflows persist up to 1 Gyr after a starburst in recently quenched galaxies at z > 1}",
      journal = {\mnras},
         year = 2024,
        month = dec,
       volume = {535},
       number = {2},
        pages = {1684-1692},
          doi = {10.1093/mnras/stae2463},
archivePrefix = {arXiv},
       eprint = {2411.00102},
 primaryClass = {astro-ph.GA},
       adsurl = {https://ui.adsabs.harvard.edu/abs/2024MNRAS.535.1684T}
}

@article{ata_predicted_2022,
	title = {Predicted future fate of {COSMOS} galaxy protoclusters over 11 {Gyr} with constrained simulations},
	volume = {6},
	issn = {2397-3366},
	url = {https://ui.adsabs.harvard.edu/abs/2022NatAs...6..857A},
	doi = {10.1038/s41550-022-01693-0},
	urldate = {2025-07-16},
	journal = {Nature Astronomy},
	author = {Ata, Metin and Lee, Khee-Gan and Vecchia, Claudio Dalla and Kitaura, Francisco-Shu and Cucciati, Olga and Lemaux, Brian C. and Kashino, Daichi and Müller, Thomas},
	month = jun,
	year = {2022},
	note = {ADS Bibcode: 2022NatAs...6..857A},
	pages = {857--865},
}

@article{kauffmann_quantitative_2014,
	title = {Quantitative constraints on starburst cycles in galaxies with stellar masses in the range 108–1010 {M}⊙},
	volume = {441},
	issn = {0035-8711},
	url = {https://doi.org/10.1093/mnras/stu752},
	doi = {10.1093/mnras/stu752},
	number = {3},
	urldate = {2025-07-15},
	journal = {Monthly Notices of the Royal Astronomical Society},
	author = {Kauffmann, Guinevere},
	month = jul,
	year = {2014},
	pages = {2717--2724},
}

@article{kurinchi-vendhan_origin_2024,
	title = {On the origin of star formation quenching in massive galaxies at z ≳ 3 in the cosmological simulations {IllustrisTNG}},
	volume = {534},
	issn = {0035-8711},
	url = {https://ui.adsabs.harvard.edu/abs/2024MNRAS.534.3974K},
	doi = {10.1093/mnras/stae2297},
	urldate = {2025-07-15},
	journal = {Monthly Notices of the Royal Astronomical Society},
	author = {Kurinchi-Vendhan, Shalini and Farcy, Marion and Hirschmann, Michaela and Valentino, Francesco},
	month = nov,
	year = {2024},
	note = {Publisher: OUP
ADS Bibcode: 2024MNRAS.534.3974K},
	pages = {3974--3988},
}

@article{donnari_quenched_2021,
	title = {Quenched fractions in the {IllustrisTNG} simulations: the roles of {AGN} feedback, environment, and pre-processing},
	volume = {500},
	issn = {0035-8711},
	shorttitle = {Quenched fractions in the {IllustrisTNG} simulations},
	url = {https://doi.org/10.1093/mnras/staa3006},
	doi = {10.1093/mnras/staa3006},
	number = {3},
	urldate = {2025-07-15},
	journal = {Monthly Notices of the Royal Astronomical Society},
	author = {Donnari, Martina and Pillepich, Annalisa and Joshi, Gandhali D and Nelson, Dylan and Genel, Shy and Marinacci, Federico and Rodriguez-Gomez, Vicente and Pakmor, Rüdiger and Torrey, Paul and Vogelsberger, Mark and Hernquist, Lars},
	month = jan,
	year = {2021},
	pages = {4004--4024},
}

@article{edward_stellar_2024,
	title = {The stellar mass function of quiescent galaxies in 2 {\textless} z {\textless} 2.5 protoclusters},
	volume = {527},
	issn = {0035-8711},
	url = {https://ui.adsabs.harvard.edu/abs/2024MNRAS.527.8598E},
	doi = {10.1093/mnras/stad3751},
	urldate = {2025-07-15},
	journal = {Monthly Notices of the Royal Astronomical Society},
	author = {Edward, Adit H. and Balogh, Michael L. and Bahé, Yannick M. and Cooper, M. C. and Hatch, Nina A. and Marchioni, Justin and Muzzin, Adam and Noble, Allison and Rudnick, Gregory H. and Vulcani, Benedetta and Wilson, Gillian and De Lucia, Gabriella and Demarco, Ricardo and Forrest, Ben and Hirschmann, Michaela and Castignani, Gianluca and Cerulo, Pierluigi and Finn, Rose A. and Hewitt, Guillaume and Jablonka, Pascale and Kodama, Tadayuki and Maurogordato, Sophie and Nantais, Julie and Xie, Lizhi},
	month = jan,
	year = {2024},
	note = {Publisher: OUP
ADS Bibcode: 2024MNRAS.527.8598E},
	pages = {8598--8617},
}

@article{mclure_sizes_2013,
	title = {The sizes, masses and specific star formation rates of massive galaxies at 1.3 {\textless} z {\textless} 1.5: strong evidence in favour of evolution via minor mergers},
	volume = {428},
	issn = {0035-8711},
	shorttitle = {The sizes, masses and specific star formation rates of massive galaxies at 1.3 {\textless} z {\textless} 1.5},
	url = {https://ui.adsabs.harvard.edu/abs/2013MNRAS.428.1088M},
	doi = {10.1093/mnras/sts092},
	urldate = {2025-07-03},
	journal = {Monthly Notices of the Royal Astronomical Society},
	author = {McLure, R. J. and Pearce, H. J. and Dunlop, J. S. and Cirasuolo, M. and Curtis-Lake, E. and Bruce, V. A. and Caputi, K. I. and Almaini, O. and Bonfield, D. G. and Bradshaw, E. J. and Buitrago, F. and Chuter, R. and Foucaud, S. and Hartley, W. G. and Jarvis, M. J.},
	month = jan,
	year = {2013},
	note = {Publisher: OUP
ADS Bibcode: 2013MNRAS.428.1088M},
	pages = {1088--1106},
}

@article{bradshaw_high-velocity_2013,
	title = {High-velocity outflows from young star-forming galaxies in the {UKIDSS} {Ultra}-{Deep} {Survey}},
	volume = {433},
	issn = {0035-8711},
	url = {https://ui.adsabs.harvard.edu/abs/2013MNRAS.433..194B},
	doi = {10.1093/mnras/stt715},
	urldate = {2025-07-03},
	journal = {Monthly Notices of the Royal Astronomical Society},
	author = {Bradshaw, E. J. and Almaini, O. and Hartley, W. G. and Smith, K. T. and Conselice, C. J. and Dunlop, J. S. and Simpson, C. and Chuter, R. W. and Cirasuolo, M. and Foucaud, S. and McLure, R. J. and Mortlock, A. and Pearce, H.},
	month = jul,
	year = {2013},
	note = {Publisher: OUP
ADS Bibcode: 2013MNRAS.433..194B},
	pages = {194--208},
}

@article{almaini_no_2025,
	title = {No evidence for excess {AGN} activity in recently quenched massive galaxies at cosmic noon},
	volume = {539},
	issn = {0035-8711},
	url = {https://doi.org/10.1093/mnras/staf659},
	doi = {10.1093/mnras/staf659},
	number = {4},
	urldate = {2025-07-03},
	journal = {Monthly Notices of the Royal Astronomical Society},
	author = {Almaini, Omar and Wild, Vivienne and Maltby, David and Taylor, Elizabeth and Rowlands, Kate and de Lisle, Thomas and Alatalo, Katherine and Harrold, Jimi and Hewitt, Guillaume and Patil, Pallavi and Skarbinski, Maya},
	month = jun,
	year = {2025},
	pages = {3568--3581},
}

@article{carnall_jwst_2024,
	title = {The {JWST} {EXCELS} survey: too much, too young, too fast? {Ultra}-massive quiescent galaxies at 3 {\textless} z {\textless} 5},
	volume = {534},
	issn = {0035-8711},
	shorttitle = {The {JWST} {EXCELS} survey},
	url = {https://ui.adsabs.harvard.edu/abs/2024MNRAS.534..325C},
	doi = {10.1093/mnras/stae2092},
	urldate = {2025-06-04},
	journal = {Monthly Notices of the Royal Astronomical Society},
	author = {Carnall, A. C. and Cullen, F. and McLure, R. J. and McLeod, D. J. and Begley, R. and Donnan, C. T. and Dunlop, J. S. and Shapley, A. E. and Rowlands, K. and Almaini, O. and Arellano-Córdova, K. Z. and Barrufet, L. and Cimatti, A. and Ellis, R. S. and Grogin, N. A. and Hamadouche, M. L. and Illingworth, G. D. and Koekemoer, A. M. and Leung, H. -H. and Lovell, C. C. and Pérez-González, P. G. and Santini, P. and Stanton, T. M. and Wild, V.},
	month = oct,
	year = {2024},
	note = {Publisher: OUP
ADS Bibcode: 2024MNRAS.534..325C},
	pages = {325--348},
}

@article{parzen_estimation_1962,
	title = {On {Estimation} of a {Probability} {Density} {Function} and {Mode}},
	volume = {33},
	issn = {0003-4851, 2168-8990},
	url = {https://projecteuclid.org/journals/annals-of-mathematical-statistics/volume-33/issue-3/On-Estimation-of-a-Probability-Density-Function-and-Mode/10.1214/aoms/1177704472.full},
	doi = {10.1214/aoms/1177704472},
	number = {3},
	urldate = {2025-06-03},
	journal = {The Annals of Mathematical Statistics},
	author = {Parzen, Emanuel},
	month = sep,
	year = {1962},
	note = {Publisher: Institute of Mathematical Statistics},
	pages = {1065--1076},
}

@article{rosenblatt_remarks_1956,
	title = {Remarks on {Some} {Nonparametric} {Estimates} of a {Density} {Function}},
	volume = {27},
	issn = {0003-4851, 2168-8990},
	url = {https://projecteuclid.org/journals/annals-of-mathematical-statistics/volume-27/issue-3/Remarks-on-Some-Nonparametric-Estimates-of-a-Density-Function/10.1214/aoms/1177728190.full},
	doi = {10.1214/aoms/1177728190},
	number = {3},
	urldate = {2025-06-03},
	journal = {The Annals of Mathematical Statistics},
	author = {Rosenblatt, Murray},
	month = sep,
	year = {1956},
	note = {Publisher: Institute of Mathematical Statistics},
	pages = {832--837},
}

@article{gully_insights_2025,
	title = {Insights into environmental quenching at z ∼ 1: an enhancement of faint, low-mass passive galaxies in clusters},
	volume = {539},
	issn = {0035-8711},
	shorttitle = {Insights into environmental quenching at z ∼ 1},
	url = {https://doi.org/10.1093/mnras/staf635},
	doi = {10.1093/mnras/staf635},
	number = {4},
	urldate = {2025-06-03},
	journal = {Monthly Notices of the Royal Astronomical Society},
	author = {Gully, Harry and Hatch, Nina and Ahad, Syeda Lammim and Bahé, Yannick and Balogh, Michael and Baxter, Devontae C and Cerulo, Pierluigi and Cooper, M C and Demarco, Ricardo and Forrest, Ben and Rescigno, Umberto and Rudnick, Gregory and Vulcani, Benedetta and Wilson, Gillian},
	month = jun,
	year = {2025},
	pages = {3058--3076},
}

@article{yan_origin_2006,
	title = {On the {Origin} of [{O} {II}] {Emission} in {Red}-{Sequence} and {Poststarburst} {Galaxies}},
	volume = {648},
	issn = {0004-637X},
	url = {https://iopscience.iop.org/article/10.1086/505629/meta},
	doi = {10.1086/505629},
	language = {en},
	number = {1},
	urldate = {2025-06-02},
	journal = {The Astrophysical Journal},
	author = {Yan, Renbin and Newman, Jeffrey A. and Faber, S. M. and Konidaris, Nicholas and Koo, David and Davis, Marc},
	month = sep,
	year = {2006},
	note = {Publisher: IOP Publishing},
	pages = {281},
}

@article{muller_multivariate_1999,
	title = {Multivariate {Boundary} {Kernels} and a {Continuous} {Least} {Squares} {Principle}},
	volume = {61},
	issn = {1369-7412},
	url = {https://doi.org/10.1111/1467-9868.00186},
	doi = {10.1111/1467-9868.00186},
	number = {2},
	urldate = {2025-04-16},
	journal = {Journal of the Royal Statistical Society Series B: Statistical Methodology},
	author = {Müller, H. G. and Stadtmüller, U.},
	month = jul,
	year = {1999},
	pages = {439--458},
}

@article{socolovsky_enhancement_2018,
	title = {The enhancement of rapidly quenched galaxies in distant clusters at 0.5 {\textless} z {\textless} 1.0},
	volume = {476},
	issn = {0035-8711},
	url = {https://ui.adsabs.harvard.edu/abs/2018MNRAS.476.1242S},
	doi = {10.1093/mnras/sty312},
	urldate = {2025-04-01},
	journal = {Monthly Notices of the Royal Astronomical Society},
	author = {Socolovsky, Miguel and Almaini, Omar and Hatch, Nina A. and Wild, Vivienne and Maltby, David T. and Hartley, William G. and Simpson, Chris},
	month = may,
	year = {2018},
	note = {Publisher: OUP
ADS Bibcode: 2018MNRAS.476.1242S},
	pages = {1242--1257},
}

@article{bezanson_massive_2013,
	title = {Massive and {Newly} {Dead}: {Discovery} of a {Significant} {Population} of {Galaxies} with {High}-velocity {Dispersions} and {Strong} {Balmer} {Lines} at z {\textasciitilde} 1.5 from {Deep} {Keck} {Spectra} and {HST}/{WFC3} {Imaging}},
	volume = {764},
	issn = {0004-637X},
	shorttitle = {Massive and {Newly} {Dead}},
	url = {https://ui.adsabs.harvard.edu/abs/2013ApJ...764L...8B},
	doi = {10.1088/2041-8205/764/1/L8},
	urldate = {2025-04-01},
	journal = {The Astrophysical Journal},
	author = {Bezanson, Rachel and van Dokkum, Pieter and van de Sande, Jesse and Franx, Marijn and Kriek, Mariska},
	month = feb,
	year = {2013},
	note = {Publisher: IOP
ADS Bibcode: 2013ApJ...764L...8B},
	pages = {L8},
}

@article{williams_morphology_2017,
	title = {Morphology {Dependence} of {Stellar} {Age} in {Quenched} {Galaxies} at {Redshift} ∼1.2: {Massive} {Compact} {Galaxies} {Are} {Older} than {More} {Extended} {Ones}},
	volume = {838},
	issn = {0004-637X},
	shorttitle = {Morphology {Dependence} of {Stellar} {Age} in {Quenched} {Galaxies} at {Redshift} ∼1.2},
	url = {https://ui.adsabs.harvard.edu/abs/2017ApJ...838...94W},
	doi = {10.3847/1538-4357/aa662f},
	urldate = {2025-04-01},
	journal = {The Astrophysical Journal},
	author = {Williams, Christina C. and Giavalisco, Mauro and Bezanson, Rachel and Cappelluti, Nico and Cassata, Paolo and Liu, Teng and Lee, Bomee and Tundo, Elena and Vanzella, Eros},
	month = apr,
	year = {2017},
	note = {Publisher: IOP
ADS Bibcode: 2017ApJ...838...94W},
	pages = {94},
}

@article{farouki_computer_1981,
	title = {Computer simulations of environmental influences on galaxy evolution in dense clusters. {II} - {Rapid} tidal encounters},
	volume = {243},
	issn = {0004-637X},
	url = {https://ui.adsabs.harvard.edu/abs/1981ApJ...243...32F},
	doi = {10.1086/158563},
	urldate = {2025-04-01},
	journal = {The Astrophysical Journal},
	author = {Farouki, R. and Shapiro, S. L.},
	month = jan,
	year = {1981},
	note = {Publisher: IOP
ADS Bibcode: 1981ApJ...243...32F},
	pages = {32--41},
}

@article{larson_evolution_1980,
	title = {The evolution of disk galaxies and the origin of {S0} galaxies},
	volume = {237},
	issn = {0004-637X},
	url = {https://ui.adsabs.harvard.edu/abs/1980ApJ...237..692L},
	doi = {10.1086/157917},
	urldate = {2025-04-01},
	journal = {The Astrophysical Journal},
	author = {Larson, R. B. and Tinsley, B. M. and Caldwell, C. N.},
	month = may,
	year = {1980},
	note = {Publisher: IOP
ADS Bibcode: 1980ApJ...237..692L},
	pages = {692--707},
}

@article{gunn_infall_1972,
	title = {On the {Infall} of {Matter} {Into} {Clusters} of {Galaxies} and {Some} {Effects} on {Their} {Evolution}},
	volume = {176},
	issn = {0004-637X},
	url = {https://ui.adsabs.harvard.edu/abs/1972ApJ...176....1G},
	doi = {10.1086/151605},
	urldate = {2025-04-01},
	journal = {The Astrophysical Journal},
	author = {Gunn, James E. and Gott, III, J. Richard},
	month = aug,
	year = {1972},
	note = {Publisher: IOP
ADS Bibcode: 1972ApJ...176....1G},
	pages = {1},
}

@article{belli_kmos3d_2017,
	title = {{KMOS3D} {Reveals} {Low}-level {Star} {Formation} {Activity} in {Massive} {Quiescent} {Galaxies} at 0.7 {\textless} z {\textless} 2.7},
	volume = {841},
	issn = {0004-637X},
	url = {https://ui.adsabs.harvard.edu/abs/2017ApJ...841L...6B},
	doi = {10.3847/2041-8213/aa70e5},
	urldate = {2025-04-01},
	journal = {The Astrophysical Journal},
	author = {Belli, Sirio and Genzel, Reinhard and Förster Schreiber, Natascha M. and Wisnioski, Emily and Wilman, David J. and Wuyts, Stijn and Mendel, J. Trevor and Beifiori, Alessandra and Bender, Ralf and Brammer, Gabriel B. and Burkert, Andreas and Chan, Jeffrey and Davies, Rebecca L. and Davies, Ric and Fabricius, Maximilian and Fossati, Matteo and Galametz, Audrey and Lang, Philipp and Lutz, Dieter and Momcheva, Ivelina G. and Nelson, Erica J. and Saglia, Roberto P. and Tacconi, Linda J. and Tadaki, Ken-ichi and Übler, Hannah and van Dokkum, Pieter},
	month = may,
	year = {2017},
	note = {Publisher: IOP
ADS Bibcode: 2017ApJ...841L...6B},
	pages = {L6},
}

@article{treu_assembly_2005,
	title = {The {Assembly} {History} of {Field} {Spheroidals}: {Evolution} of {Mass}-to-{Light} {Ratios} and {Signatures} of {Recent} {Star} {Formation}},
	volume = {633},
	issn = {0004-637X},
	shorttitle = {The {Assembly} {History} of {Field} {Spheroidals}},
	url = {https://ui.adsabs.harvard.edu/abs/2005ApJ...633..174T},
	doi = {10.1086/444585},
	urldate = {2025-04-01},
	journal = {The Astrophysical Journal},
	author = {Treu, Tommaso and Ellis, Richard S. and Liao, Ting X. and van Dokkum, Pieter G. and Tozzi, Paolo and Coil, Alison and Newman, Jeffrey and Cooper, Michael C. and Davis, Marc},
	month = nov,
	year = {2005},
	note = {Publisher: IOP
ADS Bibcode: 2005ApJ...633..174T},
	pages = {174--197},
}

@article{taylor_galaxy_2015,
	title = {Galaxy {And} {Mass} {Assembly} ({GAMA}): deconstructing bimodality - {I}. {Red} ones and blue ones},
	volume = {446},
	issn = {0035-8711},
	shorttitle = {Galaxy {And} {Mass} {Assembly} ({GAMA})},
	url = {https://ui.adsabs.harvard.edu/abs/2015MNRAS.446.2144T},
	doi = {10.1093/mnras/stu1900},
	urldate = {2025-04-01},
	journal = {Monthly Notices of the Royal Astronomical Society},
	author = {Taylor, Edward N. and Hopkins, Andrew M. and Baldry, Ivan K. and Bland-Hawthorn, Joss and Brown, Michael J. I. and Colless, Matthew and Driver, Simon and Norberg, Peder and Robotham, Aaron S. G. and Alpaslan, Mehmet and Brough, Sarah and Cluver, Michelle E. and Gunawardhana, Madusha and Kelvin, Lee S. and Liske, Jochen and Conselice, Christopher J. and Croom, Scott and Foster, Caroline and Jarrett, Thomas H. and Lara-Lopez, Maritza and Loveday, Jon},
	month = jan,
	year = {2015},
	note = {Publisher: OUP
ADS Bibcode: 2015MNRAS.446.2144T},
	pages = {2144--2185},
}

@article{baldry_quantifying_2004,
	title = {Quantifying the {Bimodal} {Color}-{Magnitude} {Distribution} of {Galaxies}},
	volume = {600},
	issn = {0004-637X},
	url = {https://ui.adsabs.harvard.edu/abs/2004ApJ...600..681B},
	doi = {10.1086/380092},
	urldate = {2025-04-01},
	journal = {The Astrophysical Journal},
	author = {Baldry, Ivan K. and Glazebrook, Karl and Brinkmann, Jon and Ivezi\'{c}, \v{Z}eljko and Lupton, Robert H. and Nichol, Robert C. and Szalay, Alexander S.},
	month = jan,
	year = {2004},
	note = {Publisher: IOP
ADS Bibcode: 2004ApJ...600..681B},
	pages = {681--694},
}

@article{ellison_low_2025,
	title = {Low redshift post-starburst galaxies host abundant {HI} reservoirs},
	volume = {8},
	issn = {2565-6120},
	url = {https://ui.adsabs.harvard.edu/abs/2025OJAp....8E..87E},
	doi = {10.33232/001c.141941},
	urldate = {2026-04-29},
	journal = {The Open Journal of Astrophysics},
	author = {Ellison, Sara and Huang, Qifeng and Yang, Dong and Wang, Jing and Wild, Vivienne and Rasmussen, Ben and Jimenez-Donaire, Maria-Jesus and Rowlands, Kate and Wilkinson, Scott and Brown, Toby and Leung, Ho-Hin},
	month = jul,
	year = {2025},
	note = {ADS Bibcode: 2025OJAp....8E..87E},
	pages = {87},
}

@article{goto_catalogue_2007,
	title = {A catalogue of local {E}+{A} (post-starburst) galaxies selected from the {Sloan} {Digital} {Sky} {Survey} {Data} {Release} 5},
	volume = {381},
	issn = {0035-8711},
	url = {https://doi.org/10.1111/j.1365-2966.2007.12227.x},
	doi = {10.1111/j.1365-2966.2007.12227.x},
	number = {1},
	urldate = {2025-01-09},
	journal = {Monthly Notices of the Royal Astronomical Society},
	author = {Goto, Tomotsugu},
	month = oct,
	year = {2007},
	pages = {187--193},
}

@article{pentericci_vandels_2018,
	title = {The {VANDELS} {ESO} public spectroscopic survey: {Observations} and first data release},
	volume = {616},
	copyright = {© ESO 2018},
	issn = {0004-6361, 1432-0746},
	shorttitle = {The {VANDELS} {ESO} public spectroscopic survey},
	url = {https://www.aanda.org/articles/aa/abs/2018/08/aa33047-18/aa33047-18.html},
	doi = {10.1051/0004-6361/201833047},
	language = {en},
	urldate = {2024-11-29},
	journal = {Astronomy \& Astrophysics},
	author = {Pentericci, L. and McLure, R. J. and Garilli, B. and Cucciati, O. and Franzetti, P. and Iovino, A. and Amorin, R. and Bolzonella, M. and Bongiorno, A. and Carnall, A. C. and Castellano, M. and Cimatti, A. and Cirasuolo, M. and Cullen, F. and Barros, S. De and Dunlop, J. S. and Elbaz, D. and Finkelstein, S. L. and Fontana, A. and Fontanot, F. and Fumana, M. and Gargiulo, A. and Guaita, L. and Hartley, W. G. and Jarvis, M. J. and Juneau, S. and Karman, W. and Maccagni, D. and Marchi, F. and Marmol-Queralto, E. and Nandra, K. and Pompei, E. and Pozzetti, L. and Scodeggio, M. and Sommariva, V. and Talia, M. and Almaini, O. and Balestra, I. and Bardelli, S. and Bell, E. F. and Bourne, N. and Bowler, R. a. A. and Brusa, M. and Buitrago, F. and Caputi, K. I. and Cassata, P. and Charlot, S. and Citro, A. and Cresci, G. and Cristiani, S. and Curtis-Lake, E. and Dickinson, M. and Fazio, G. G. and Ferguson, H. C. and Fiore, F. and Franco, M. and Fynbo, J. P. U. and Galametz, A. and Georgakakis, A. and Giavalisco, M. and Grazian, A. and Hathi, N. P. and Jung, I. and Kim, S. and Koekemoer, A. M. and Khusanova, Y. and Fèvre, O. Le and Lotz, J. M. and Mannucci, F. and Maltby, D. T. and Matsuoka, K. and McLeod, D. J. and Mendez-Hernandez, H. and Mendez-Abreu, J. and Mignoli, M. and Moresco, M. and Mortlock, A. and Nonino, M. and Pannella, M. and Papovich, C. and Popesso, P. and Rosario, D. P. and Salvato, M. and Santini, P. and Schaerer, D. and Schreiber, C. and Stark, D. P. and Tasca, L. a. M. and Thomas, R. and Treu, T. and Vanzella, E. and Wild, V. and Williams, C. C. and Zamorani, G. and Zucca, E.},
	month = aug,
	year = {2018},
	note = {Publisher: EDP Sciences},
	pages = {A174},
}

@article{mclure_vandels_2018,
	title = {The {VANDELS} {ESO} public spectroscopic survey},
	volume = {479},
	issn = {0035-8711},
	url = {https://doi.org/10.1093/mnras/sty1213},
	doi = {10.1093/mnras/sty1213},
	number = {1},
	urldate = {2024-11-29},
	journal = {Monthly Notices of the Royal Astronomical Society},
	author = {McLure, R J and Pentericci, L and Cimatti, A and Dunlop, J S and Elbaz, D and Fontana, A and Nandra, K and Amorin, R and Bolzonella, M and Bongiorno, A and Carnall, A C and Castellano, M and Cirasuolo, M and Cucciati, O and Cullen, F and De Barros, S and Finkelstein, S L and Fontanot, F and Franzetti, P and Fumana, M and Gargiulo, A and Garilli, B and Guaita, L and Hartley, W G and Iovino, A and Jarvis, M J and Juneau, S and Karman, W and Maccagni, D and Marchi, F and Mármol-Queraltó, E and Pompei, E and Pozzetti, L and Scodeggio, M and Sommariva, V and Talia, M and Almaini, O and Balestra, I and Bardelli, S and Bell, E F and Bourne, N and Bowler, R A A and Brusa, M and Buitrago, F and Caputi, K I and Cassata, P and Charlot, S and Citro, A and Cresci, G and Cristiani, S and Curtis-Lake, E and Dickinson, M and Fazio, G G and Ferguson, H C and Fiore, F and Franco, M and Fynbo, J P U and Galametz, A and Georgakakis, A and Giavalisco, M and Grazian, A and Hathi, N P and Jung, I and Kim, S and Koekemoer, A M and Khusanova, Y and Le Fèvre, O and Lotz, J M and Mannucci, F and Maltby, D T and Matsuoka, K and McLeod, D J and Mendez-Hernandez, H and Mendez-Abreu, J and Mignoli, M and Moresco, M and Mortlock, A and Nonino, M and Pannella, M and Papovich, C and Popesso, P and Rosario, D P and Salvato, M and Santini, P and Schaerer, D and Schreiber, C and Stark, D P and Tasca, L A M and Thomas, R and Treu, T and Vanzella, E and Wild, V and Williams, C C and Zamorani, G and Zucca, E},
	month = sep,
	year = {2018},
	pages = {25--42},
}

@article{dunlop_primer_2021,
	title = {{PRIMER}: {Public} {Release} {IMaging} for {Extragalactic} {Research}},
	shorttitle = {{PRIMER}},
	url = {https://ui.adsabs.harvard.edu/abs/2021jwst.prop.1837D},
	urldate = {2024-11-27},
	journal = {JWST Proposal. Cycle 1},
	author = {Dunlop, James S. and Abraham, Roberto G. and Ashby, Matthew L. N. and Bagley, Micaela and Best, Philip N. and Bongiorno, Angela and Bouwens, Rychard and Bowler, Rebecca A. A. and Brammer, Gabriel and Bremer, Malcolm and Calabro', Antonello and Carnall, Adam and Castellano, Marco and Cirasuolo, Michele and Conselice, Christopher and Cullen, Fergus and Dave, Romeel and Dayal, Pratika and Dekel, Avishai and Dickinson, Mark and Duncan, Kenneth James and Elbaz, David and Ellis, Richard S. and Ferguson, Harry C. and Ferrara, Andrea and Finkelstein, Steven L. and Fontana, Adriano and Furlanetto, Steven and Fynbo, Johan P. U. and Gallerani, Simona and Gardner, Jonathan P. and Giavalisco, Mauro and Grazian, Andrea and Grogin, Norman and Harikane, Yuichi and Hopkins, Philip F. and Ilbert, Olivier and Illingworth, Garth D. and Juneau, Stephanie and Jung, Intae and Kartaltepe, Jeyhan and Kassin, Susan and Kauffmann, Olivier Benjamin and Khochfar, Sadegh and Kirkpatrick, Allison and Kocevski, Dale D. and Koekemoer, Anton M. and Labbe, Ivo and Laporte, Nicolas and Larson, Rebecca L. and Lucas, Ray A. and Magee, Daniel K. and Mason, Charlotte and McCracken, Henry Joy and McLeod, Derek and McLure, Ross and Merlin, Emiliano and Mesinger, Andrei and Milvang-Jensen, Bo and Newman, Jeffrey Allen and Oesch, Pascal and Ouchi, Masami and Pacifici, Camilla and Papovich, Casey and Peacock, John and Peeples, Molly and Pentericci, Laura and Perez-Gonzalez, Pablo G. and Pirzkal, Norbert and Pope, Alexandra and Pye, John P. and Reddy, Naveen A. and Robertson, Brant and Salvato, Mara and Santini, Paola and Schaerer, Daniel and Shapley, Alice E. and Simons, Raymond and Smit, Renske and Smith, Britton D. and Snyder, Greg and Somerville, Rachel S. and Stanway, Elizabeth R. and Stefanon, Mauro and Tasca, Lidia and Tikkanen, Tuomo and Tresse, Laurence and Trump, Jonathan R. and Whitaker, Katherine E. and Wilkins, Stephen Matthew and Wright, Gillian and Wyithe, J. Stuart B. and van Dokkum, Pieter and van der Werf, Paul},
	month = mar,
	year = {2021},
	note = {ADS Bibcode: 2021jwst.prop.1837D},
	pages = {1837},
}

@article{galametz_growing_2018,
	title = {Growing up in a megalopolis: environmental effects on galaxy evolution in a supercluster at z ∼ 0.65 in {UKIDSS} {UDS}},
	volume = {475},
	issn = {0035-8711},
	shorttitle = {Growing up in a megalopolis},
	url = {https://doi.org/10.1093/mnras/sty095},
	doi = {10.1093/mnras/sty095},
	number = {3},
	urldate = {2024-08-06},
	journal = {Monthly Notices of the Royal Astronomical Society},
	author = {Galametz, Audrey and Pentericci, Laura and Castellano, Marco and Mendel, Trevor and Hartley, Will G and Fossati, Matteo and Finoguenov, Alexis and Almaini, Omar and Beifiori, Alessandra and Fontana, Adriano and Grazian, Andrea and Scodeggio, Marco and Kocevski, Dale D},
	month = apr,
	year = {2018},
	pages = {4148--4169},
}

@article{papovich_spitzer-selected_2010,
	title = {A {SPITZER}-{SELECTED} {GALAXY} {CLUSTER} {AT} z = 1.62*},
	volume = {716},
	issn = {0004-637X},
	url = {https://dx.doi.org/10.1088/0004-637X/716/2/1503},
	doi = {10.1088/0004-637X/716/2/1503},
	language = {en},
	number = {2},
	urldate = {2024-08-06},
	journal = {The Astrophysical Journal},
	author = {Papovich, C. and Momcheva, I. and Willmer, C. N. A. and Finkelstein, K. D. and Finkelstein, S. L. and Tran, K.-V. and Brodwin, M. and Dunlop, J. S. and Farrah, D. and Khan, S. A. and Lotz, J. and McCarthy, P. and McLure, R. J. and Rieke, M. and Rudnick, G. and Sivanandam, S. and Pacaud, F. and Pierre, M.},
	month = jun,
	year = {2010},
	note = {Publisher: The American Astronomical Society},
	pages = {1503},
}

@article{zhang_desi_2024,
	title = {{DESI} {Massive} {Poststarburst} {Galaxies} at z ∼ 1.2 {Have} {Compact} {Structures} and {Dense} {Cores}},
	volume = {976},
	issn = {0004-637X},
	url = {https://ui.adsabs.harvard.edu/abs/2024ApJ...976...36Z},
	doi = {10.3847/1538-4357/ad7c45},
	urldate = {2025-11-17},
	journal = {The Astrophysical Journal},
	author = {Zhang, Yunchong and Setton, David J. and Price, Sedona H. and Bezanson, Rachel and Khullar, Gourav and Newman, Jeffrey A. and Aguilar, Jessica Nicole and Ahlen, Steven and Andrews, Brett H. and Brooks, David and Claybaugh, Todd and de la Macorra, Axel and Dey, Biprateep and Doel, Peter and Gaztañaga, Enrique and Gontcho A Gontcho, Satya and Greene, Jenny E. and Juneau, Stephanie and Kehoe, Robert and Kisner, Theodore and Kriek, Mariska and Leja, Joel and Manera, Marc and Meisner, Aaron and Miquel, Ramon and Moustakas, John and Prada, Francisco and Rossi, Graziano and Sanchez, Eusebio and Schubnell, Michael and Siudek, Małgorzata and Spilker, Justin and Sprayberry, David and Suess, Katherine A. and Tarlé, Gregory and Zou, Hu and {DESI Collaboration}},
	month = nov,
	year = {2024}
}

@article{grogin_candels_2011,
	title = {{CANDELS}: {THE} {COSMIC} {ASSEMBLY} {NEAR}-{INFRARED} {DEEP} {EXTRAGALACTIC} {LEGACY} {SURVEY}},
	volume = {197},
	issn = {0067-0049},
	shorttitle = {{CANDELS}},
	url = {https://dx.doi.org/10.1088/0067-0049/197/2/35},
	doi = {10.1088/0067-0049/197/2/35},
	language = {en},
	number = {2},
	urldate = {2024-07-31},
	journal = {The Astrophysical Journal Supplement Series},
	author = {Grogin, Norman A. and Kocevski, Dale D. and Faber, S. M. and Ferguson, Henry C. and Koekemoer, Anton M. and Riess, Adam G. and Acquaviva, Viviana and Alexander, David M. and Almaini, Omar and Ashby, Matthew L. N. and Barden, Marco and Bell, Eric F. and Bournaud, Frédéric and Brown, Thomas M. and Caputi, Karina I. and Casertano, Stefano and Cassata, Paolo and Castellano, Marco and Challis, Peter and Chary, Ranga-Ram and Cheung, Edmond and Cirasuolo, Michele and Conselice, Christopher J. and Cooray, Asantha Roshan and Croton, Darren J. and Daddi, Emanuele and Dahlen, Tomas and Davé, Romeel and Mello, Duília F. de and Dekel, Avishai and Dickinson, Mark and Dolch, Timothy and Donley, Jennifer L. and Dunlop, James S. and Dutton, Aaron A. and Elbaz, David and Fazio, Giovanni G. and Filippenko, Alexei V. and Finkelstein, Steven L. and Fontana, Adriano and Gardner, Jonathan P. and Garnavich, Peter M. and Gawiser, Eric and Giavalisco, Mauro and Grazian, Andrea and Guo, Yicheng and Hathi, Nimish P. and Häussler, Boris and Hopkins, Philip F. and Huang, Jia-Sheng and Huang, Kuang-Han and Jha, Saurabh W. and Kartaltepe, Jeyhan S. and Kirshner, Robert P. and Koo, David C. and Lai, Kamson and Lee, Kyoung-Soo and Li, Weidong and Lotz, Jennifer M. and Lucas, Ray A. and Madau, Piero and McCarthy, Patrick J. and McGrath, Elizabeth J. and McIntosh, Daniel H. and McLure, Ross J. and Mobasher, Bahram and Moustakas, Leonidas A. and Mozena, Mark and Nandra, Kirpal and Newman, Jeffrey A. and Niemi, Sami-Matias and Noeske, Kai G. and Papovich, Casey J. and Pentericci, Laura and Pope, Alexandra and Primack, Joel R. and Rajan, Abhijith and Ravindranath, Swara and Reddy, Naveen A. and Renzini, Alvio and Rix, Hans-Walter and Robaina, Aday R. and Rodney, Steven A. and Rosario, David J. and Rosati, Piero and Salimbeni, Sara and Scarlata, Claudia and Siana, Brian and Simard, Luc and Smidt, Joseph and Somerville, Rachel S. and Spinrad, Hyron and Straughn, Amber N. and Strolger, Louis-Gregory and Telford, Olivia and Teplitz, Harry I. and Trump, Jonathan R. and Wel, Arjen van der and Villforth, Carolin and Wechsler, Risa H. and Weiner, Benjamin J. and Wiklind, Tommy and Wild, Vivienne and Wilson, Grant and Wuyts, Stijn and Yan, Hao-Jing and Yun, Min S.},
	month = dec,
	year = {2011},
	note = {Publisher: The American Astronomical Society},
	pages = {35},
}

@article{koekemoer_candels_2011,
	title = {{CANDELS}: {The} {Cosmic} {Assembly} {Near}-infrared {Deep} {Extragalactic} {Legacy} {Survey}—{The} {Hubble} {Space} {Telescope} {Observations}, {Imaging} {Data} {Products}, and {Mosaics}},
	volume = {197},
	issn = {0067-0049},
	shorttitle = {{CANDELS}},
	url = {https://ui.adsabs.harvard.edu/abs/2011ApJS..197...36K},
	doi = {10.1088/0067-0049/197/2/36},
	urldate = {2024-07-31},
	journal = {The Astrophysical Journal Supplement Series},
	author = {Koekemoer, Anton M. and Faber, S. M. and Ferguson, Henry C. and Grogin, Norman A. and Kocevski, Dale D. and Koo, David C. and Lai, Kamson and Lotz, Jennifer M. and Lucas, Ray A. and McGrath, Elizabeth J. and Ogaz, Sara and Rajan, Abhijith and Riess, Adam G. and Rodney, Steve A. and Strolger, Louis and Casertano, Stefano and Castellano, Marco and Dahlen, Tomas and Dickinson, Mark and Dolch, Timothy and Fontana, Adriano and Giavalisco, Mauro and Grazian, Andrea and Guo, Yicheng and Hathi, Nimish P. and Huang, Kuang-Han and van der Wel, Arjen and Yan, Hao-Jing and Acquaviva, Viviana and Alexander, David M. and Almaini, Omar and Ashby, Matthew L. N. and Barden, Marco and Bell, Eric F. and Bournaud, Frédéric and Brown, Thomas M. and Caputi, Karina I. and Cassata, Paolo and Challis, Peter J. and Chary, Ranga-Ram and Cheung, Edmond and Cirasuolo, Michele and Conselice, Christopher J. and Roshan Cooray, Asantha and Croton, Darren J. and Daddi, Emanuele and Davé, Romeel and de Mello, Duilia F. and de Ravel, Loic and Dekel, Avishai and Donley, Jennifer L. and Dunlop, James S. and Dutton, Aaron A. and Elbaz, David and Fazio, Giovanni G. and Filippenko, Alexei V. and Finkelstein, Steven L. and Frazer, Chris and Gardner, Jonathan P. and Garnavich, Peter M. and Gawiser, Eric and Gruetzbauch, Ruth and Hartley, Will G. and Häussler, Boris and Herrington, Jessica and Hopkins, Philip F. and Huang, Jia-Sheng and Jha, Saurabh W. and Johnson, Andrew and Kartaltepe, Jeyhan S. and Khostovan, Ali A. and Kirshner, Robert P. and Lani, Caterina and Lee, Kyoung-Soo and Li, Weidong and Madau, Piero and McCarthy, Patrick J. and McIntosh, Daniel H. and McLure, Ross J. and McPartland, Conor and Mobasher, Bahram and Moreira, Heidi and Mortlock, Alice and Moustakas, Leonidas A. and Mozena, Mark and Nandra, Kirpal and Newman, Jeffrey A. and Nielsen, Jennifer L. and Niemi, Sami and Noeske, Kai G. and Papovich, Casey J. and Pentericci, Laura and Pope, Alexandra and Primack, Joel R. and Ravindranath, Swara and Reddy, Naveen A. and Renzini, Alvio and Rix, Hans-Walter and Robaina, Aday R. and Rosario, David J. and Rosati, Piero and Salimbeni, Sara and Scarlata, Claudia and Siana, Brian and Simard, Luc and Smidt, Joseph and Snyder, Diana and Somerville, Rachel S. and Spinrad, Hyron and Straughn, Amber N. and Telford, Olivia and Teplitz, Harry I. and Trump, Jonathan R. and Vargas, Carlos and Villforth, Carolin and Wagner, Cory R. and Wandro, Pat and Wechsler, Risa H. and Weiner, Benjamin J. and Wiklind, Tommy and Wild, Vivienne and Wilson, Grant and Wuyts, Stijn and Yun, Min S.},
	month = dec,
	year = {2011},
	note = {Publisher: IOP
ADS Bibcode: 2011ApJS..197...36K},
	pages = {36},
}

@article{brammer_eazy_2008,
	title = {{EAZY}: {A} {Fast}, {Public} {Photometric} {Redshift} {Code}},
	volume = {686},
	issn = {0004-637X},
	shorttitle = {{EAZY}},
	url = {https://iopscience.iop.org/article/10.1086/591786/meta},
	doi = {10.1086/591786},
	language = {en},
	number = {2},
	urldate = {2024-07-31},
	journal = {The Astrophysical Journal},
	author = {Brammer, Gabriel B. and Dokkum, Pieter G. van and Coppi, Paolo},
	month = oct,
	year = {2008},
	note = {Publisher: IOP Publishing},
	pages = {1503},
}

@article{baldry_galaxy_2012,
	title = {Galaxy {And} {Mass} {Assembly} ({GAMA}): the galaxy stellar mass function at z {\textless} 0.06},
	volume = {421},
	issn = {0035-8711},
	shorttitle = {Galaxy {And} {Mass} {Assembly} ({GAMA})},
	url = {https://ui.adsabs.harvard.edu/abs/2012MNRAS.421..621B},
	doi = {10.1111/j.1365-2966.2012.20340.x},
	urldate = {2024-07-30},
	journal = {Monthly Notices of the Royal Astronomical Society},
	author = {Baldry, I. K. and Driver, S. P. and Loveday, J. and Taylor, E. N. and Kelvin, L. S. and Liske, J. and Norberg, P. and Robotham, A. S. G. and Brough, S. and Hopkins, A. M. and Bamford, S. P. and Peacock, J. A. and Bland-Hawthorn, J. and Conselice, C. J. and Croom, S. M. and Jones, D. H. and Parkinson, H. R. and Popescu, C. C. and Prescott, M. and Sharp, R. G. and Tuffs, R. J.},
	month = mar,
	year = {2012},
	note = {Publisher: OUP
ADS Bibcode: 2012MNRAS.421..621B},
	pages = {621--634},
}

@article{tomczak_galaxy_2014,
	title = {Galaxy {Stellar} {Mass} {Functions} from {ZFOURGE}/{CANDELS}: {An} {Excess} of {Low}-mass {Galaxies} since z = 2 and the {Rapid} {Buildup} of {Quiescent} {Galaxies}},
	volume = {783},
	issn = {0004-637X},
	shorttitle = {Galaxy {Stellar} {Mass} {Functions} from {ZFOURGE}/{CANDELS}},
	url = {https://ui.adsabs.harvard.edu/abs/2014ApJ...783...85T},
	doi = {10.1088/0004-637X/783/2/85},
	urldate = {2024-07-30},
	journal = {The Astrophysical Journal},
	author = {Tomczak, Adam R. and Quadri, Ryan F. and Tran, Kim-Vy H. and Labbé, Ivo and Straatman, Caroline M. S. and Papovich, Casey and Glazebrook, Karl and Allen, Rebecca and Brammer, Gabriel B. and Kacprzak, Glenn G. and Kawinwanichakij, Lalitwadee and Kelson, Daniel D. and McCarthy, Patrick J. and Mehrtens, Nicola and Monson, Andrew J. and Persson, S. Eric and Spitler, Lee R. and Tilvi, Vithal and van Dokkum, Pieter},
	month = mar,
	year = {2014},
	note = {Publisher: IOP
ADS Bibcode: 2014ApJ...783...85T},
	pages = {85},
}

@article{whitaker_large_2012,
	title = {A {LARGE} {POPULATION} {OF} {MASSIVE} {COMPACT} {POST}-{STARBURST} {GALAXIES} {AT} z {\textgreater} 1: {IMPLICATIONS} {FOR} {THE} {SIZE} {EVOLUTION} {AND} {QUENCHING} {MECHANISM} {OF} {QUIESCENT} {GALAXIES}},
	volume = {745},
	issn = {0004-637X},
	shorttitle = {A {LARGE} {POPULATION} {OF} {MASSIVE} {COMPACT} {POST}-{STARBURST} {GALAXIES} {AT} z {\textgreater} 1},
	url = {https://dx.doi.org/10.1088/0004-637X/745/2/179},
	doi = {10.1088/0004-637X/745/2/179},
	language = {en},
	number = {2},
	urldate = {2024-07-29},
	journal = {The Astrophysical Journal},
	author = {Whitaker, Katherine E. and Kriek, Mariska and Dokkum, Pieter G. van and Bezanson, Rachel and Brammer, Gabriel and Franx, Marijn and Labbé, Ivo},
	month = jan,
	year = {2012},
	note = {Publisher: The American Astronomical Society},
	pages = {179},
}

@article{furusawa_subaruxmm-newton_2008,
	title = {The {Subaru}/{XMM}-{Newton} {Deep} {Survey} ({SXDS}). {II}. {Optical} {Imaging} and {Photometric} {Catalogs}*},
	volume = {176},
	issn = {0067-0049},
	url = {https://iopscience.iop.org/article/10.1086/527321/meta},
	doi = {10.1086/527321},
	language = {en},
	number = {1},
	urldate = {2024-07-29},
	journal = {The Astrophysical Journal Supplement Series},
	author = {Furusawa, Hisanori and Kosugi, George and Akiyama, Masayuki and Takata, Tadafumi and Sekiguchi, Kazuhiro and Tanaka, Ichi and Iwata, Ikuru and Kajisawa, Masaru and Yasuda, Naoki and Doi, Mamoru and Ouchi, Masami and Simpson, Chris and Shimasaku, Kazuhiro and Yamada, Toru and Furusawa, Junko and Morokuma, Tomoki and Ishida, Catherine M. and Aoki, Kentaro and Fuse, Tetsuharu and Imanishi, Masatoshi and Iye, Masanori and Karoji, Hiroshi and Kobayashi, Naoto and Kodama, Tadayuki and Komiyama, Yutaka and Maeda, Yoshitomo and Miyazaki, Satoshi and Mizumoto, Yoshihiko and Nakata, Fumiaki and Noumaru, Jun’ichi and Ogasawara, Ryusuke and Okamura, Sadanori and Saito, Tomoki and Sasaki, Toshiyuki and Ueda, Yoshihiro and Yoshida, Michitoshi},
	month = may,
	year = {2008},
	note = {Publisher: IOP Publishing},
	pages = {1},
}

@article{bruzual_stellar_2003,
	title = {Stellar population synthesis at the resolution of 2003},
	volume = {344},
	issn = {0035-8711},
	url = {https://ui.adsabs.harvard.edu/abs/2003MNRAS.344.1000B},
	doi = {10.1046/j.1365-8711.2003.06897.x},
	urldate = {2024-07-26},
	journal = {Monthly Notices of the Royal Astronomical Society},
	author = {Bruzual, G. and Charlot, S.},
	month = oct,
	year = {2003},
	note = {Publisher: OUP
ADS Bibcode: 2003MNRAS.344.1000B},
	pages = {1000--1028},
}

@article{belli_mosfire_2019,
	title = {{MOSFIRE} {Spectroscopy} of {Quiescent} {Galaxies} at 1.5 {\textless} z {\textless} 2.5. {II}. {Star} {Formation} {Histories} and {Galaxy} {Quenching}},
	volume = {874},
	issn = {0004-637X},
	url = {https://dx.doi.org/10.3847/1538-4357/ab07af},
	doi = {10.3847/1538-4357/ab07af},
	language = {en},
	number = {1},
	urldate = {2024-07-17},
	journal = {The Astrophysical Journal},
	author = {Belli, Sirio and Newman, Andrew B. and Ellis, Richard S.},
	month = mar,
	year = {2019},
	note = {Publisher: The American Astronomical Society},
	pages = {17},
}

@article{pozzetti_zcosmos_2010,
	title = {{zCOSMOS} - 10k-bright spectroscopic sample. {The} bimodality in the galaxy stellar mass function: exploring its evolution with redshift},
	volume = {523},
	issn = {0004-6361},
	shorttitle = {{zCOSMOS} - 10k-bright spectroscopic sample. {The} bimodality in the galaxy stellar mass function},
	url = {https://ui.adsabs.harvard.edu/abs/2010A&A...523A..13P},
	doi = {10.1051/0004-6361/200913020},
	urldate = {2024-06-14},
	journal = {Astronomy and Astrophysics},
	author = {Pozzetti, L. and Bolzonella, M. and Zucca, E. and Zamorani, G. and Lilly, S. and Renzini, A. and Moresco, M. and Mignoli, M. and Cassata, P. and Tasca, L. and Lamareille, F. and Maier, C. and Meneux, B. and Halliday, C. and Oesch, P. and Vergani, D. and Caputi, K. and Kovač, K. and Cimatti, A. and Cucciati, O. and Iovino, A. and Peng, Y. and Carollo, M. and Contini, T. and Kneib, J. -P. and Le Févre, O. and Mainieri, V. and Scodeggio, M. and Bardelli, S. and Bongiorno, A. and Coppa, G. and de la Torre, S. and de Ravel, L. and Franzetti, P. and Garilli, B. and Kampczyk, P. and Knobel, C. and Le Borgne, J. -F. and Le Brun, V. and Pellò, R. and Perez Montero, E. and Ricciardelli, E. and Silverman, J. D. and Tanaka, M. and Tresse, L. and Abbas, U. and Bottini, D. and Cappi, A. and Guzzo, L. and Koekemoer, A. M. and Leauthaud, A. and Maccagni, D. and Marinoni, C. and McCracken, H. J. and Memeo, P. and Porciani, C. and Scaramella, R. and Scarlata, C. and Scoville, N.},
	month = nov,
	year = {2010},
	note = {ADS Bibcode: 2010A\&A...523A..13P},
	pages = {A13},
}

@article{santini_stellar_2022,
	title = {The {Stellar} {Mass} {Function} in {CANDELS} and {Frontier} {Fields}: {The} {Buildup} of {Low}-mass {Passive} {Galaxies} since z   3},
	volume = {940},
	issn = {0004-637X},
	shorttitle = {The {Stellar} {Mass} {Function} in {CANDELS} and {Frontier} {Fields}},
	url = {https://ui.adsabs.harvard.edu/abs/2022ApJ...940..135S},
	doi = {10.3847/1538-4357/ac9a48},
	urldate = {2024-06-12},
	journal = {The Astrophysical Journal},
	author = {Santini, Paola and Castellano, Marco and Fontana, Adriano and Fortuni, Flaminia and Menci, Nicola and Merlin, Emiliano and Pagul, Amanda and Testa, Vincenzo and Calabrò, Antonello and Paris, Diego and Pentericci, Laura},
	month = dec,
	year = {2022},
	note = {Publisher: IOP
ADS Bibcode: 2022ApJ...940..135S},
	pages = {135},
}

@article{mcleod_evolution_2021,
	title = {The evolution of the galaxy stellar-mass function over the last 12 billion years from a combination of ground-based and {HST} surveys},
	volume = {503},
	issn = {0035-8711},
	url = {https://ui.adsabs.harvard.edu/abs/2021MNRAS.503.4413M},
	doi = {10.1093/mnras/stab731},
	urldate = {2024-04-29},
	journal = {Monthly Notices of the Royal Astronomical Society},
	author = {McLeod, D. J. and McLure, R. J. and Dunlop, J. S. and Cullen, F. and Carnall, A. C. and Duncan, K.},
	month = may,
	year = {2021},
	note = {Publisher: OUP
ADS Bibcode: 2021MNRAS.503.4413M},
	pages = {4413--4435},
}

@article{taylor_role_2023,
	title = {The role of mass and environment in the build-up of the quenched galaxy population since cosmic noon},
	volume = {522},
	issn = {0035-8711},
	url = {https://doi.org/10.1093/mnras/stad1098},
	doi = {10.1093/mnras/stad1098},
	number = {2},
	urldate = {2024-02-06},
	journal = {Monthly Notices of the Royal Astronomical Society},
	author = {Taylor, Elizabeth and Almaini, Omar and Merrifield, Michael and Maltby, David and Wild, Vivienne and Hartley, William G and Rowlands, Kate},
	month = jun,
	year = {2023},
	pages = {2297--2306},
}

@article{wild_star_2020,
	title = {The star formation histories of z ∼ 1 post-starburst galaxies},
	volume = {494},
	issn = {0035-8711},
	url = {https://doi.org/10.1093/mnras/staa674},
	doi = {10.1093/mnras/staa674},
	number = {1},
	urldate = {2023-09-21},
	journal = {Monthly Notices of the Royal Astronomical Society},
	author = {Wild, Vivienne and Taj Aldeen, Laith and Carnall, Adam and Maltby, David and Almaini, Omar and Werle, Ariel and Wilkinson, Aaron and Rowlands, Kate and Bolzonella, Micol and Castellano, Marco and Gargiulo, Adriana and McLure, Ross and Pentericci, Laura and Pozzetti, Lucia},
	month = may,
	year = {2020},
	pages = {529--548},
}

@article{dekel_formation_2009,
	title = {Formation of {Massive} {Galaxies} at {High} {Redshift}: {Cold} {Streams}, {Clumpy} {Disks}, and {Compact} {Spheroids}},
	volume = {703},
	issn = {0004-637X},
	shorttitle = {Formation of {Massive} {Galaxies} at {High} {Redshift}},
	url = {https://dx.doi.org/10.1088/0004-637X/703/1/785},
	doi = {10.1088/0004-637X/703/1/785},
	language = {en},
	number = {1},
	urldate = {2023-07-28},
	journal = {The Astrophysical Journal},
	author = {Dekel, Avishai and Sari, Re'em and Ceverino, Daniel},
	month = sep,
	year = {2009},
	note = {Publisher: The American Astronomical Society},
	pages = {785},
}

@article{muzzin_evolution_2013,
	title = {The {Evolution} of the {S}℡lar {Mass} {Functions} of {Star}-{Forming} and {Quiescent} {Galaxies} to {Z} = 4 from the {Cosmos}/{Ultravista} {Survey}*},
	volume = {777},
	issn = {0004-637X},
	url = {https://dx.doi.org/10.1088/0004-637X/777/1/18},
	doi = {10.1088/0004-637X/777/1/18},
	language = {en},
	number = {1},
	urldate = {2023-07-31},
	journal = {The Astrophysical Journal},
	author = {Muzzin, Adam and Marchesini, Danilo and Stefanon, Mauro and Franx, Marijn and McCracken, Henry J. and Milvang-Jensen, Bo and Dunlop, James S. and Fynbo, J. P. U. and Brammer, Gabriel and Labbé, Ivo and Dokkum, Pieter G. van},
	month = oct,
	year = {2013},
	note = {Publisher: The American Astronomical Society},
	pages = {18},
}

@article{wu_fast_2018,
	title = {Fast and {Slow} {Paths} to {Quiescence}: {Ages} and {Sizes} of 400 {Quiescent} {Galaxies} from the {LEGA}-{C} {Survey}},
	volume = {868},
	issn = {0004-637X},
	shorttitle = {Fast and {Slow} {Paths} to {Quiescence}},
	url = {https://dx.doi.org/10.3847/1538-4357/aae822},
	doi = {10.3847/1538-4357/aae822},
	language = {en},
	number = {1},
	urldate = {2023-07-31},
	journal = {The Astrophysical Journal},
	author = {Wu, Po-Feng and Wel, Arjen van der and Bezanson, Rachel and Gallazzi, Anna and Pacifici, Camilla and Straatman, Caroline M. S. and Barišić, Ivana and Bell, Eric F. and Chauke, Priscilla and Houdt, Josha van and Franx, Marijn and Muzzin, Adam and Sobral, David and Wild, Vivienne},
	month = nov,
	year = {2018},
	note = {Publisher: The American Astronomical Society},
	pages = {37},
}

@article{maltby_structure_2018,
	title = {The structure of post-starburst galaxies at 0.5 {\textless} z {\textless} 2: evidence for two distinct quenching routes at different epochs},
	volume = {480},
	issn = {0035-8711},
	shorttitle = {The structure of post-starburst galaxies at 0.5 {\textless} z {\textless} 2},
	url = {https://doi.org/10.1093/mnras/sty1794},
	doi = {10.1093/mnras/sty1794},
	number = {1},
	urldate = {2023-07-31},
	journal = {Monthly Notices of the Royal Astronomical Society},
	author = {Maltby, David T and Almaini, Omar and Wild, Vivienne and Hatch, Nina A and Hartley, William G and Simpson, Chris and Rowlands, Kate and Socolovsky, Miguel},
	month = oct,
	year = {2018},
	pages = {381--401},
}

@article{maltby_identification_2016,
	title = {The identification of post-starburst galaxies at z ∼ 1 using multiwavelength photometry: a spectroscopic verification},
	volume = {459},
	issn = {1745-3925},
	shorttitle = {The identification of post-starburst galaxies at z ∼ 1 using multiwavelength photometry},
	url = {https://doi.org/10.1093/mnrasl/slw057},
	doi = {10.1093/mnrasl/slw057},
	number = {1},
	urldate = {2023-07-31},
	journal = {Monthly Notices of the Royal Astronomical Society: Letters},
	author = {Maltby, David T. and Almaini, Omar and Wild, Vivienne and Hatch, Nina A. and Hartley, William G. and Simpson, Chris and McLure, Ross J. and Dunlop, James and Rowlands, Kate and Cirasuolo, Michele},
	month = jun,
	year = {2016},
	pages = {L114--L118},
}

@article{french_discovery_2015,
	title = {{DISCOVERY} {OF} {LARGE} {MOLECULAR} {GAS} {RESERVOIRS} {IN} {POST}-{STARBURST} {GALAXIES}},
	volume = {801},
	issn = {0004-637X},
	url = {https://dx.doi.org/10.1088/0004-637X/801/1/1},
	doi = {10.1088/0004-637X/801/1/1},
	language = {en},
	number = {1},
	urldate = {2023-07-31},
	journal = {The Astrophysical Journal},
	author = {French, K. Decker and Yang, Yujin and Zabludoff, Ann and Narayanan, Desika and Shirley, Yancy and Walter, Fabian and Smith, John-David and Tremonti, Christy A.},
	month = feb,
	year = {2015},
	note = {Publisher: The American Astronomical Society},
	pages = {1},
}

@article{rowlands_evolution_2015,
	title = {The evolution of the cold interstellar medium in galaxies following a starburst★},
	volume = {448},
	issn = {0035-8711},
	url = {https://doi.org/10.1093/mnras/stu2714},
	doi = {10.1093/mnras/stu2714},
	number = {1},
	urldate = {2023-07-31},
	journal = {Monthly Notices of the Royal Astronomical Society},
	author = {Rowlands, K. and Wild, V. and Nesvadba, N. and Sibthorpe, B. and Mortier, A. and Lehnert, M. and da Cunha, E.},
	month = mar,
	year = {2015},
	pages = {258--279},
}

@article{zwaan_cold_2013,
	title = {The cold gas content of post-starburst galaxies},
	volume = {432},
	issn = {0035-8711},
	url = {https://doi.org/10.1093/mnras/stt496},
	doi = {10.1093/mnras/stt496},
	number = {1},
	urldate = {2023-07-31},
	journal = {Monthly Notices of the Royal Astronomical Society},
	author = {Zwaan, Martin A. and Kuntschner, Harald and Pracy, Michael B. and Couch, Warrick J.},
	month = jun,
	year = {2013},
	pages = {492--499},
}

@article{pawlik_shape_2016,
	title = {Shape asymmetry: a morphological indicator for automatic detection of galaxies in the post-coalescence merger stages},
	volume = {456},
	issn = {0035-8711},
	shorttitle = {Shape asymmetry},
	url = {https://doi.org/10.1093/mnras/stv2878},
	doi = {10.1093/mnras/stv2878},
	number = {3},
	urldate = {2023-07-31},
	journal = {Monthly Notices of the Royal Astronomical Society},
	author = {Pawlik, M. M. and Wild, V. and Walcher, C. J. and Johansson, P. H. and Villforth, C. and Rowlands, K. and Mendez-Abreu, J. and Hewlett, T.},
	month = mar,
	year = {2016},
	pages = {3032--3052},
}

@article{yang_detailed_2008,
	title = {The {Detailed} {Evolution} of {E}+{A} {Galaxies} into {Early} {Types}*},
	volume = {688},
	issn = {0004-637X},
	url = {https://iopscience.iop.org/article/10.1086/591656/meta},
	doi = {10.1086/591656},
	language = {en},
	number = {2},
	urldate = {2023-07-31},
	journal = {The Astrophysical Journal},
	author = {Yang, Yujin and Zabludoff, Ann I. and Zaritsky, Dennis and Mihos, J. Christopher},
	month = dec,
	year = {2008},
	note = {Publisher: IOP Publishing},
	pages = {945},
}

@article{blake_2df_2004,
	title = {The {2dF} {Galaxy} {Redshift} {Survey}: the local {E}+{A} galaxy population},
	volume = {355},
	issn = {0035-8711},
	shorttitle = {The {2dF} {Galaxy} {Redshift} {Survey}},
	url = {https://doi.org/10.1111/j.1365-2966.2004.08351.x},
	doi = {10.1111/j.1365-2966.2004.08351.x},
	number = {3},
	urldate = {2023-07-31},
	journal = {Monthly Notices of the Royal Astronomical Society},
	author = {Blake, Chris and Pracy, Michael B. and Couch, Warrick J. and Bekki, Kenji and Lewis, Ian and Glazebrook, Karl and Baldry, Ivan K. and Baugh, Carlton M. and Bland-Hawthorn, Joss and Bridges, Terry and Cannon, Russell and Cole, Shaun and Colless, Matthew and Collins, Chris and Dalton, Gavin and De Propris, Roberto and Driver, Simon P. and Efstathiou, George and Ellis, Richard S. and Frenk, Carlos S. and Jackson, Carole and Lahav, Ofer and Lumsden, Stuart and Maddox, Steve and Madgwick, Darren and Norberg, Peder and Peacock, John A. and Peterson, Bruce A. and Sutherland, Will and Taylor, Keith},
	month = dec,
	year = {2004},
	pages = {713--727},
}

@article{zabludoff_environment_1996,
	title = {The {Environment} of ``{E}+{A}'' {Galaxies}},
	volume = {466},
	issn = {0004-637X},
	url = {https://ui.adsabs.harvard.edu/abs/1996ApJ...466..104Z},
	doi = {10.1086/177495},
	urldate = {2023-07-31},
	journal = {The Astrophysical Journal},
	author = {Zabludoff, Ann I. and Zaritsky, Dennis and Lin, Huan and Tucker, Douglas and Hashimoto, Yasuhiro and Shectman, Stephen A. and Oemler, Augustus and Kirshner, Robert P.},
	month = jul,
	year = {1996},
	note = {ADS Bibcode: 1996ApJ...466..104Z},
	pages = {104},
}

@article{goto_266_2005,
	title = {266 {E}+{A} galaxies selected from the {Sloan} {Digital} {Sky} {Survey} {Data} {Release} 2: the origin of {E}+{A} galaxies},
	volume = {357},
	issn = {0035-8711},
	shorttitle = {266 {E}+{A} galaxies selected from the {Sloan} {Digital} {Sky} {Survey} {Data} {Release} 2},
	url = {https://doi.org/10.1111/j.1365-2966.2005.08701.x},
	doi = {10.1111/j.1365-2966.2005.08701.x},
	number = {3},
	urldate = {2023-07-31},
	journal = {Monthly Notices of the Royal Astronomical Society},
	author = {Goto, Tomotsugu},
	month = mar,
	year = {2005},
	pages = {937--944},
}

@article{jarvis_vista_2013,
	title = {The {VISTA} {Deep} {Extragalactic} {Observations} ({VIDEO}) survey},
	volume = {428},
	issn = {0035-8711},
	url = {https://ui.adsabs.harvard.edu/abs/2013MNRAS.428.1281J},
	doi = {10.1093/mnras/sts118},
	urldate = {2023-07-31},
	journal = {Monthly Notices of the Royal Astronomical Society},
	author = {Jarvis, Matt J. and Bonfield, D. G. and Bruce, V. A. and Geach, J. E. and McAlpine, K. and McLure, R. J. and González-Solares, E. and Irwin, M. and Lewis, J. and Yoldas, A. Kupcu and Andreon, S. and Cross, N. J. G. and Emerson, J. P. and Dalton, G. and Dunlop, J. S. and Hodgkin, S. T. and Le, Fèvre O. and Karouzos, M. and Meisenheimer, K. and Oliver, S. and Rawlings, S. and Simpson, C. and Smail, I. and Smith, D. J. B. and Sullivan, M. and Sutherland, W. and White, S. V. and Zwart, J. T. L.},
	month = jan,
	year = {2013},
	note = {ADS Bibcode: 2013MNRAS.428.1281J},
	pages = {1281--1295},
}

@article{vergani_kgalaxies_2010,
	title = {K+a galaxies in the {zCOSMOS} survey - {Physical} properties of systems in their post-starburst phase},
	volume = {509},
	copyright = {© ESO, 2010},
	issn = {0004-6361, 1432-0746},
	url = {https://www.aanda.org/articles/aa/abs/2010/01/aa12802-09/aa12802-09.html},
	doi = {10.1051/0004-6361/200912802},
	language = {en},
	urldate = {2023-07-31},
	journal = {Astronomy \& Astrophysics},
	author = {Vergani, D. and Zamorani, G. and Lilly, S. and Lamareille, F. and Halliday, C. and Scodeggio, M. and Vignali, C. and Ciliegi, P. and Bolzonella, M. and Bondi, M. and Kovač, K. and Knobel, C. and Zucca, E. and Caputi, K. and Pozzetti, L. and Bardelli, S. and Mignoli, M. and Iovino, A. and Carollo, C. M. and Contini, T. and Kneib, J.-P. and Fèvre, O. Le and Mainieri, V. and Renzini, A. and Bongiorno, A. and Coppa, G. and Cucciati, O. and Torre, S. de la and Ravel, L. de and Franzetti, P. and Garilli, B. and Kampczyk, P. and Borgne, J.-F. Le and Brun, V. Le and Maier, C. and Pello, R. and Peng, Y. and Montero, E. Perez and Ricciardelli, E. and Silverman, J. D. and Tanaka, M. and Tasca, L. and Tresse, L. and Abbas, U. and Bottini, D. and Cappi, A. and Cassata, P. and Cimatti, A. and Guzzo, L. and Koekemoer, A. M. and Leauthaud, A. and Maccagni, D. and Marinoni, C. and McCracken, H. J. and Memeo, P. and Meneux, B. and Oesch, P. and Porciani, C. and Scaramella, R. and Capak, P. and Sanders, D. and Scoville, N. and Taniguchi, Y.},
	month = jan,
	year = {2010},
	note = {Publisher: EDP Sciences},
	pages = {A42},
}

@article{wild_post-starburst_2009,
	title = {Post-starburst galaxies: more than just an interesting curiosity},
	volume = {395},
	issn = {0035-8711},
	shorttitle = {Post-starburst galaxies},
	url = {https://doi.org/10.1111/j.1365-2966.2009.14537.x},
	doi = {10.1111/j.1365-2966.2009.14537.x},
	number = {1},
	urldate = {2023-07-31},
	journal = {Monthly Notices of the Royal Astronomical Society},
	author = {Wild, Vivienne and Walcher, C. Jakob and Johansson, Peter H. and Tresse, Laurence and Charlot, Stéphane and Pollo, Agnieszka and Le Fèvre, Olivier and De Ravel, Loic},
	month = may,
	year = {2009},
	pages = {144--159},
}

@article{faber_galaxy_2007,
	title = {Galaxy {Luminosity} {Functions} to z 1 from {DEEP2} and {COMBO}-17: {Implications} for {Red} {Galaxy} {Formation}*},
	volume = {665},
	issn = {0004-637X},
	shorttitle = {Galaxy {Luminosity} {Functions} to z 1 from {DEEP2} and {COMBO}-17},
	url = {https://dx.doi.org/10.1086/519294},
	doi = {10.1086/519294},
	language = {en},
	number = {1},
	urldate = {2023-07-31},
	journal = {The Astrophysical Journal},
	author = {Faber, S. M. and Willmer, C. N. A. and Wolf, C. and Koo, D. C. and Weiner, B. J. and Newman, J. A. and Im, M. and Coil, A. L. and Conroy, C. and Cooper, M. C. and Davis, M. and Finkbeiner, D. P. and Gerke, B. F. and Gebhardt, K. and Groth, E. J. and Guhathakurta, P. and Harker, J. and Kaiser, N. and Kassin, S. and Kleinheinrich, M. and Konidaris, N. P. and Kron, R. G. and Lin, L. and Luppino, G. and Madgwick, D. S. and Meisenheimer, K. and Noeske, K. G. and Phillips, A. C. and Sarajedini, V. L. and Schiavon, R. P. and Simard, L. and Szalay, A. S. and Vogt, N. P. and Yan, R.},
	month = aug,
	year = {2007},
	pages = {265},
}

@article{bell_nearly_2004,
	title = {Nearly 5000 {Distant} {Early}-{Type} {Galaxies} in {COMBO}-17: {A} {Red} {Sequence} and {Its} {Evolution} since z 1},
	volume = {608},
	issn = {0004-637X},
	shorttitle = {Nearly 5000 {Distant} {Early}-{Type} {Galaxies} in {COMBO}-17},
	url = {https://dx.doi.org/10.1086/420778},
	doi = {10.1086/420778},
	language = {en},
	number = {2},
	urldate = {2023-07-31},
	journal = {The Astrophysical Journal},
	author = {Bell, Eric F. and Wolf, Christian and Meisenheimer, Klaus and Rix, Hans-Walter and Borch, Andrea and Dye, Simon and Kleinheinrich, Martina and Wisotzki, Lutz and McIntosh, Daniel H.},
	month = jun,
	year = {2004},
	pages = {752},
}

@article{dressler_spectroscopy_1983,
	title = {Spectroscopy of galaxies in distant clusters. {II}. {The} population of the {3C} 295 cluster.},
	volume = {270},
	issn = {0004-637X},
	url = {https://ui.adsabs.harvard.edu/abs/1983ApJ...270....7D},
	doi = {10.1086/161093},
	urldate = {2023-07-31},
	journal = {The Astrophysical Journal},
	author = {Dressler, A. and Gunn, J. E.},
	month = jul,
	year = {1983},
	note = {ADS Bibcode: 1983ApJ...270....7D},
	pages = {7--19},
}

@article{wilkinson_starburst_2021,
	title = {From starburst to quiescence: post-starburst galaxies and their large-scale clustering over cosmic time},
	volume = {504},
	issn = {0035-8711},
	shorttitle = {From starburst to quiescence},
	url = {https://doi.org/10.1093/mnras/stab965},
	doi = {10.1093/mnras/stab965},
	number = {3},
	urldate = {2023-07-28},
	journal = {Monthly Notices of the Royal Astronomical Society},
	author = {Wilkinson, Aaron and Almaini, Omar and Wild, Vivienne and Maltby, David and Hartley, William G and Simpson, Chris and Rowlands, Kate},
	month = jul,
	year = {2021},
	pages = {4533--4550},
}

@article{lawrence_ukirt_2007,
	title = {The {UKIRT} {Infrared} {Deep} {Sky} {Survey} ({UKIDSS})},
	volume = {379},
	issn = {0035-8711},
	url = {https://doi.org/10.1111/j.1365-2966.2007.12040.x},
	doi = {10.1111/j.1365-2966.2007.12040.x},
	number = {4},
	urldate = {2023-07-28},
	journal = {Monthly Notices of the Royal Astronomical Society},
	author = {Lawrence, A. and Warren, S. J. and Almaini, O. and Edge, A. C. and Hambly, N. C. and Jameson, R. F. and Lucas, P. and Casali, M. and Adamson, A. and Dye, S. and Emerson, J. P. and Foucaud, S. and Hewett, P. and Hirst, P. and Hodgkin, S. T. and Irwin, M. J. and Lodieu, N. and McMahon, R. G. and Simpson, C. and Smail, I. and Mortlock, D. and Folger, M.},
	month = aug,
	year = {2007},
	pages = {1599--1617},
}

@book{toomre_mergers_1977,
	title = {Mergers and {Some} {Consequences}},
	url = {https://ui.adsabs.harvard.edu/abs/1977egsp.conf..401T},
	urldate = {2023-07-28},
	author = {Toomre, Alar},
	month = jan,
	year = {1977},
	note = {Conference Name: Evolution of Galaxies and Stellar Populations
Pages: 401
ADS Bibcode: 1977egsp.conf..401T},
}

@article{thomas_environment_2010,
	title = {Environment and self-regulation in galaxy formation},
	volume = {404},
	issn = {0035-8711},
	url = {https://doi.org/10.1111/j.1365-2966.2010.16427.x},
	doi = {10.1111/j.1365-2966.2010.16427.x},
	number = {4},
	urldate = {2023-07-28},
	journal = {Monthly Notices of the Royal Astronomical Society},
	author = {Thomas, Daniel and Maraston, Claudia and Schawinski, Kevin and Sarzi, Marc and Silk, Joseph},
	month = jun,
	year = {2010},
	pages = {1775--1789},
}

@article{hopkins_black_2005,
	title = {Black {Holes} in {Galaxy} {Mergers}: {Evolution} of {Quasars}},
	volume = {630},
	issn = {0004-637X, 1538-4357},
	shorttitle = {Black {Holes} in {Galaxy} {Mergers}},
	url = {https://iopscience.iop.org/article/10.1086/432438},
	doi = {10.1086/432438},
	language = {en},
	number = {2},
	urldate = {2023-07-28},
	journal = {The Astrophysical Journal},
	author = {Hopkins, Philip F. and Hernquist, Lars and Cox, Thomas J. and Di Matteo, Tiziana and Martini, Paul and Robertson, Brant and Springel, Volker},
	month = sep,
	year = {2005},
	pages = {705--715},
}

@article{strateva_color_2001,
	title = {Color {Separation} of {Galaxy} {Types} in the {Sloan} {Digital} {Sky} {Survey} {Imaging} {Data}},
	volume = {122},
	issn = {1538-3881},
	url = {https://iopscience.iop.org/article/10.1086/323301/meta},
	doi = {10.1086/323301},
	language = {en},
	number = {4},
	urldate = {2023-07-28},
	journal = {The Astronomical Journal},
	author = {Strateva, Iskra and Ivezi\'{c}, \v{Z}eljko and Knapp, Gillian R. and Narayanan, Vijay K. and Strauss, Michael A. and Gunn, James E. and Lupton, Robert H. and Schlegel, David and Bahcall, Neta A. and Brinkmann, Jon and Brunner, Robert J. and Budavári, Tamás and Csabai, István and Castander, Francisco Javier and Doi, Mamoru and Fukugita, Masataka and Győry, Zsuzsanna and Hamabe, Masaru and Hennessy, Greg and Ichikawa, Takashi and Kunszt, Peter Z. and Lamb, Don Q. and McKay, Timothy A. and Okamura, Sadanori and Racusin, Judith and Sekiguchi, Maki and Schneider, Donald P. and Shimasaku, Kazuhiro and York, Donald},
	month = oct,
	year = {2001},
	note = {Publisher: IOP Publishing},
	pages = {1861},
}

@article{wild_evolution_2016,
	title = {The evolution of post-starburst galaxies from z=2 to 0.5},
	volume = {463},
	issn = {0035-8711},
	url = {https://doi.org/10.1093/mnras/stw1996},
	doi = {10.1093/mnras/stw1996},
	number = {1},
	urldate = {2023-07-28},
	journal = {Monthly Notices of the Royal Astronomical Society},
	author = {Wild, Vivienne and Almaini, Omar and Dunlop, Jim and Simpson, Chris and Rowlands, Kate and Bowler, Rebecca and Maltby, David and McLure, Ross},
	month = nov,
	year = {2016},
	pages = {832--844},
}

@article{almaini_massive_2017,
	title = {Massive post-starburst galaxies at z {\textgreater} 1 are compact proto-spheroids},
	volume = {472},
	issn = {0035-8711},
	url = {https://doi.org/10.1093/mnras/stx1957},
	doi = {10.1093/mnras/stx1957},
	number = {2},
	urldate = {2023-07-28},
	journal = {Monthly Notices of the Royal Astronomical Society},
	author = {Almaini, Omar and Wild, Vivienne and Maltby, David T. and Hartley, William G. and Simpson, Chris and Hatch, Nina A. and McLure, Ross J. and Dunlop, James S. and Rowlands, Kate},
	month = dec,
	year = {2017},
	pages = {1401--1412},
}

@article{maltby_high-velocity_2019,
	title = {High-velocity outflows in massive post-starburst galaxies at z {\textgreater} 1},
	volume = {489},
	issn = {0035-8711},
	url = {https://doi.org/10.1093/mnras/stz2211},
	doi = {10.1093/mnras/stz2211},
	number = {1},
	urldate = {2023-07-28},
	journal = {Monthly Notices of the Royal Astronomical Society},
	author = {Maltby, David T and Almaini, Omar and McLure, Ross J and Wild, Vivienne and Dunlop, James and Rowlands, Kate and Hartley, William G and Hatch, Nina A and Socolovsky, Miguel and Wilkinson, Aaron and Amorin, Ricardo and Bradshaw, Emma J and Carnall, Adam C and Castellano, Marco and Cimatti, Andrea and Cresci, Giovanni and Cullen, Fergus and De Barros, Stephane and Fontanot, Fabio and Garilli, Bianca and Koekemoer, Anton M and McLeod, Derek J and Pentericci, Laura and Talia, Margherita},
	month = oct,
	year = {2019},
	pages = {1139--1151},
}

@article{wild_new_2014,
	title = {A new method for classifying galaxy {SEDs} from multiwavelength photometry},
	volume = {440},
	issn = {0035-8711},
	url = {https://doi.org/10.1093/mnras/stu212},
	doi = {10.1093/mnras/stu212},
	number = {2},
	urldate = {2022-11-24},
	journal = {Monthly Notices of the Royal Astronomical Society},
	author = {Wild, Vivienne and Almaini, Omar and Cirasuolo, Michele and Dunlop, Jim and McLure, Ross and Bowler, Rebecca and Ferreira, Joao and Bradshaw, Emma and Chuter, Robert and Hartley, Will},
	month = may,
	year = {2014},
	pages = {1880--1898},
}

@article{pawlik_origins_2018,
	title = {The origins of post-starburst galaxies at z {\textless} 0.05},
	volume = {477},
	issn = {0035-8711},
	url = {https://ui.adsabs.harvard.edu/abs/2018MNRAS.477.1708P},
	doi = {10.1093/mnras/sty589},
	urldate = {2025-09-08},
	journal = {Monthly Notices of the Royal Astronomical Society},
	author = {Pawlik, M. M. and Taj Aldeen, L. and Wild, V. and Mendez-Abreu, J. and Lahén, N. and Johansson, P. H. and Jimenez, N. and Lucas, W. and Zheng, Y. and Walcher, C. J. and Rowlands, K.},
	month = jun,
	year = {2018},
}

@article{wilkinson_merger_2022,
	title = {The merger fraction of post-starburst galaxies in {UNIONS}},
	volume = {516},
	issn = {0035-8711},
	url = {https://ui.adsabs.harvard.edu/abs/2022MNRAS.516.4354W},
	doi = {10.1093/mnras/stac1962},
	urldate = {2025-09-09},
	journal = {Monthly Notices of the Royal Astronomical Society},
	author = {Wilkinson, Scott and Ellison, Sara L. and Bottrell, Connor and Bickley, Robert W. and Gwyn, Stephen and Cuillandre, Jean-Charles and Wild, Vivienne},
	month = nov,
	year = {2022}
}

@article{ellison_galaxy_2022,
	title = {Galaxy mergers can rapidly shut down star formation},
	volume = {517},
	issn = {0035-8711},
	url = {https://ui.adsabs.harvard.edu/abs/2022MNRAS.517L..92E},
	doi = {10.1093/mnrasl/slac109},
	urldate = {2025-09-09},
	journal = {Monthly Notices of the Royal Astronomical Society},
	author = {Ellison, Sara L. and Wilkinson, Scott and Woo, Joanna and Leung, Ho-Hin and Wild, Vivienne and Bickley, Robert W. and Patton, David R. and Quai, Salvatore and Gwyn, Stephen},
	month = nov,
	year = {2022}
}

@article{soto_catalog_2025,
	title = {A {Catalog} of {Post}-starburst {Galaxies} from {DESI}},
	volume = {9},
	issn = {2515-5172},
	url = {https://doi.org/10.3847/2515-5172/adeb89},
	doi = {10.3847/2515-5172/adeb89},
	language = {en},
	number = {7},
	urldate = {2025-11-17},
	journal = {Research Notes of the AAS},
	author = {Soto, Dharlenny and French, K. Decker},
	month = jul,
	year = {2025}
}

@article{muzzin_phase_2014,
	title = {{THE} {PHASE} {SPACE} {AND} {S}℡{LAR} {POPULATIONS} {OF} {CLUSTER} {GALAXIES} {AT} z ∼ 1: {SIMULTANEOUS} {CONSTRAINTS} {ON} {THE} {LOCATION} {AND} {TIMESCALE} {OF} {SA}℡{LITE} {QUENCHING}*},
	volume = {796},
	issn = {0004-637X},
	shorttitle = {{THE} {PHASE} {SPACE} {AND} {S}℡{LAR} {POPULATIONS} {OF} {CLUSTER} {GALAXIES} {AT} z ∼ 1},
	url = {https://doi.org/10.1088/0004-637X/796/1/65},
	doi = {10.1088/0004-637X/796/1/65},
	language = {en},
	number = {1},
	urldate = {2025-11-27},
	journal = {The Astrophysical Journal},
	author = {Muzzin, Adam and van der Burg, R. F. J. and McGee, Sean L. and Balogh, Michael and Franx, Marijn and Hoekstra, Henk and Hudson, Michael J. and Noble, Allison and Taranu, Dan S. and Webb, Tracy and Wilson, Gillian and Yee, H. K. C.},
	month = nov,
	year = {2014}
}

@article{mcnab_gogreen_2021,
	title = {The {GOGREEN} survey: transition galaxies and the evolution of environmental quenching},
	volume = {508},
	issn = {0035-8711},
	shorttitle = {The {GOGREEN} survey},
	url = {https://doi.org/10.1093/mnras/stab2558},
	doi = {10.1093/mnras/stab2558},
	number = {1},
	urldate = {2025-11-27},
	journal = {Monthly Notices of the Royal Astronomical Society},
	author = {McNab, Karen and Balogh, Michael L and van der Burg, Remco F J and Forestell, Anya and Webb, Kristi and Vulcani, Benedetta and Rudnick, Gregory and Muzzin, Adam and Cooper, M C and McGee, Sean and Biviano, Andrea and Cerulo, Pierluigi and Chan, Jeffrey C C and De Lucia, Gabriella and Demarco, Ricardo and Finoguenov, Alexis and Forrest, Ben and Golledge, Caelan and Jablonka, Pascale and Lidman, Chris and Nantais, Julie and Old, Lyndsay and Pintos-Castro, Irene and Poggianti, Bianca and Reeves, Andrew M M and Wilson, Gillian and Yee, Howard K C and Zaritsky, Dennis},
	month = nov,
	year = {2021},
	pages = {157--174}
}

@article{atek_star_2022,
	title = {The star formation burstiness and ionizing efficiency of low-mass galaxies},
	volume = {511},
	issn = {0035-8711},
	doi = {10.1093/mnras/stac360},
	number = {3},
	journal = {Monthly Notices of the Royal Astronomical Society},
	author = {Atek, Hakim and Furtak, Lukas J. and Oesch, Pascal and van Dokkum, Pieter and Reddy, Naveen and Contini, Thierry and Illingworth, Garth and Wilkins, Stephen},
	month = feb,
	year = {2022},
	pages = {4464--4479},
}

@article{gendron_numerical_2025,
	title = {Numerical simulations of starbursts triggered by tidal disruption in clusters: searching for observational signatures},
	volume = {541},
	issn = {0035-8711},
	shorttitle = {Numerical simulations of starbursts triggered by tidal disruption in clusters},
	url = {https://ui.adsabs.harvard.edu/abs/2025MNRAS.541.2513G},
	doi = {10.1093/mnras/staf1142},
	urldate = {2026-01-07},
	journal = {Monthly Notices of the Royal Astronomical Society},
	author = {Gendron, Valérie and Martel, Hugo},
	month = aug,
	year = {2025},
	note = {Publisher: OUP
ADS Bibcode: 2025MNRAS.541.2513G},
	pages = {2513--2539},
}

@article{stevenson_primer_2026,
	title = {{PRIMER} and {JADES} reveal an abundance of massive quiescent galaxies at 2 {\textless} z {\textless} 5},
	volume = {545},
	issn = {0035-8711},
	url = {https://ui.adsabs.harvard.edu/abs/2026MNRAS.545f2087S},
	doi = {10.1093/mnras/staf2087},
	urldate = {2026-01-09},
	journal = {Monthly Notices of the Royal Astronomical Society},
	author = {Stevenson, Struan D. and Carnall, Adam C. and Leung, Ho-Hin and Taylor, Elizabeth and Cullen, Fergus and Dunlop, James S. and McLeod, Derek J. and McLure, Ross J. and Begley, Ryan and Arellano-Córdova, Karla Z. and Barrufet, Laia and Bondestam, Cecilia and Donnan, Callum T. and Ellis, Richard S. and Grogin, Norman A. and Koekemoer, Anton M. and Liu, Feng-Yuan and Pérez-González, Pablo G. and Rowlands, Kate and Sanders, Ryan L. and Scholte, Dirk and Shapley, Alice E. and Skarbinski, Maya and Stanton, Thomas M. and Wild, Vivienne},
	month = jan,
	year = {2026},
	note = {Publisher: OUP
ADS Bibcode: 2026MNRAS.545f2087S},
	pages = {staf2087},
}

@book{Scott1992,
  author    = {Scott, David W.},
  title     = {Multivariate Density Estimation: Theory, Practice, and Visualization},
  publisher = {John Wiley \& Sons},
  address   = {New York},
  year      = {1992}
}

@article{flores_velazquez_time-scales_2021,
	title = {The time-scales probed by star formation rate indicators for realistic, bursty star formation histories from the {FIRE} simulations},
	volume = {501},
	issn = {0035-8711},
	url = {https://ui.adsabs.harvard.edu/abs/2021MNRAS.501.4812F},
	doi = {10.1093/mnras/staa3893},
	urldate = {2026-01-09},
	journal = {Monthly Notices of the Royal Astronomical Society},
	author = {Flores Velázquez, José A. and Gurvich, Alexander B. and Faucher-Giguère, Claude-André and Bullock, James S. and Starkenburg, Tjitske K. and Moreno, Jorge and Lazar, Alexandres and Mercado, Francisco J. and Stern, Jonathan and Sparre, Martin and Hayward, Christopher C. and Wetzel, Andrew and El-Badry, Kareem},
	month = mar,
	year = {2021},
	note = {Publisher: OUP
ADS Bibcode: 2021MNRAS.501.4812F},
	pages = {4812--4824},
}

@article{shen_baryon_2014,
	title = {The {Baryon} {Cycle} of {Dwarf} {Galaxies}: {Dark}, {Bursty}, {Gas}-rich {Polluters}},
	volume = {792},
	issn = {0004-637X},
	shorttitle = {The {Baryon} {Cycle} of {Dwarf} {Galaxies}},
	url = {https://ui.adsabs.harvard.edu/abs/2014ApJ...792...99S},
	doi = {10.1088/0004-637X/792/2/99},
	urldate = {2026-01-09},
	journal = {The Astrophysical Journal},
	author = {Shen, Sijing and Madau, Piero and Conroy, Charlie and Governato, Fabio and Mayer, Lucio},
	month = sep,
	year = {2014},
	note = {Publisher: IOP
ADS Bibcode: 2014ApJ...792...99S},
	pages = {99},
}

@article{harrold_role_2026,
	title = {The role of mergers and rejuvenation in the buildup of the quiescent population at cosmic noon},
	volume = {545},
	issn = {0035-8711},
	url = {https://doi.org/10.1093/mnras/staf2191},
	doi = {10.1093/mnras/staf2191},
	number = {3},
	urldate = {2026-01-12},
	journal = {Monthly Notices of the Royal Astronomical Society},
	author = {Harrold, Jimi E and Almaini, Omar and Pearce, Frazer R and Yates, Robert M and Maltby, Dave and Rowlands, Kate and Wild, Vivienne and Skarbinski, Maya and de Lisle, Thomas},
	month = jan,
	year = {2026},
	pages = {staf2191},
}

@article{dressler_imacs_2013,
	title = {The {IMACS} {Cluster} {Building} {Survey}. {II}. {Spectral} {Evolution} of {Galaxies} in the {Epoch} of {Cluster} {Assembly}},
	volume = {770},
	issn = {0004-637X},
	url = {https://ui.adsabs.harvard.edu/abs/2013ApJ...770...62D},
	doi = {10.1088/0004-637X/770/1/62},
	urldate = {2025-09-08},
	journal = {The Astrophysical Journal},
	author = {Dressler, Alan and Oemler, Jr., Augustus and Poggianti, Bianca M. and Gladders, Michael D. and Abramson, Louis and Vulcani, Benedetta},
	month = jun,
	year = {2013},
	note = {ADS Bibcode: 2013ApJ...770...62D},
	pages = {62},
}

@article{peng_mass_2010,
	title = {Mass and {Environment} as {Drivers} of {Galaxy} {Evolution} in {SDSS} and {zCOSMOS} and the {Origin} of the {Schechter} {Function}},
	volume = {721},
	issn = {0004-637X},
	url = {https://ui.adsabs.harvard.edu/abs/2010ApJ...721..193P},
	doi = {10.1088/0004-637X/721/1/193},
	urldate = {2026-04-20},
	journal = {The Astrophysical Journal},
	publisher = {IOP},
	author = {Peng, Ying-jie and Lilly, Simon J. and Kovač, Katarina and Bolzonella, Micol and Pozzetti, Lucia and Renzini, Alvio and Zamorani, Gianni and Ilbert, Olivier and Knobel, Christian and Iovino, Angela and Maier, Christian and Cucciati, Olga and Tasca, Lidia and Carollo, C. Marcella and Silverman, John and Kampczyk, Pawel and de Ravel, Loic and Sanders, David and Scoville, Nicholas and Contini, Thierry and Mainieri, Vincenzo and Scodeggio, Marco and Kneib, Jean-Paul and Le Fèvre, Olivier and Bardelli, Sandro and Bongiorno, Angela and Caputi, Karina and Coppa, Graziano and de la Torre, Sylvain and Franzetti, Paolo and Garilli, Bianca and Lamareille, Fabrice and Le Borgne, Jean-Francois and Le Brun, Vincent and Mignoli, Marco and Perez Montero, Enrique and Pello, Roser and Ricciardelli, Elena and Tanaka, Masayuki and Tresse, Laurence and Vergani, Daniela and Welikala, Niraj and Zucca, Elena and Oesch, Pascal and Abbas, Ummi and Barnes, Luke and Bordoloi, Rongmon and Bottini, Dario and Cappi, Alberto and Cassata, Paolo and Cimatti, Andrea and Fumana, Marco and Hasinger, Gunther and Koekemoer, Anton and Leauthaud, Alexei and Maccagni, Dario and Marinoni, Christian and McCracken, Henry and Memeo, Pierdomenico and Meneux, Baptiste and Nair, Preethi and Porciani, Cristiano and Presotto, Valentina and Scaramella, Roberto},
	month = sep,
	year = {2010},
	note = {ADS Bibcode: 2010ApJ...721..193P},
	pages = {193--221},
}

@article{bower_dark_2017,
	title = {The dark nemesis of galaxy formation: why hot haloes trigger black hole growth and bring star formation to an end},
	volume = {465},
	issn = {0035-8711},
	shorttitle = {The dark nemesis of galaxy formation},
	url = {https://ui.adsabs.harvard.edu/abs/2017MNRAS.465...32B},
	doi = {10.1093/mnras/stw2735},
	urldate = {2026-04-20},
	journal = {Monthly Notices of the Royal Astronomical Society},
	publisher = {OUP},
	author = {Bower, Richard G. and Schaye, Joop and Frenk, Carlos S. and Theuns, Tom and Schaller, Matthieu and Crain, Robert A. and McAlpine, Stuart},
	month = feb,
	year = {2017},
	note = {ADS Bibcode: 2017MNRAS.465...32B},
	pages = {32--44},
}





\bsp	
\label{lastpage}
\end{document}